\documentclass{aa}

\usepackage{graphicx}
\usepackage{txfonts}
\usepackage{amsmath}	
\usepackage{amssymb}	
\usepackage{xspace}
\usepackage[colorlinks=true,linkcolor=blue,citecolor=blue]{hyperref}
\usepackage{ulem}
\usepackage[section]{placeins}
\usepackage{url}
\usepackage{longtable} 
\usepackage{booktabs}
\usepackage{multirow}
\usepackage{appendix}
\usepackage{natbib}

\begin{document}

   \title{Simulating quasar microlensing light curves: High magnification events}

    \author{}
   \author{F. Neira\inst{1,2}\fnmsep\thanks{fcneirad@gmail.com}
          \and
          T. Anguita\inst{1,3}
          \and
          G. Vernardos\inst{4,5,6}
          }

   \institute{Instituto de Astrofísica, Facultad de Ciencias Exactas, Universidad Andres Bello, Av. Fernandez Concha 700, Las Condes, Santiago, Chile
              \and
              Institute of Physics, Laboratory of Astrophysics,  Ecole Polytechnique Fédérale de Lausanne (EPFL), Observatoire de Sauverny 1290 Versoix, Switzerland
              \and
              Millennium Institute of Astrophysics, Monseñor Nuncio Sotero Sanz 100, Oficina 104, 7500011 Providencia, Santiago, Chile
              \and
              The Graduate Center of the City University of New York, 365 Fifth Avenue, New York, NY 10016, USA
              \and
              Department of Astrophysics, American Museum of Natural History, Central Park West and 79th Street, NY 10024-5192, USA
              \and
              Department of Physics and Astronomy, Lehman College of the CUNY, Bronx, NY 10468, USA
             }

\date{Accepted XXX. Received YYY; in original form ZZZ}

\abstract{Quasar microlensing can be used to constrain important astrophysical properties, such as the accretion disk size and the amount of stars in the lensing galaxy.
The associated brightness variations over time, in particular high magnification events (HMEs) and caustic crossings, can yield precise constraints due to their strong dependence on the relative projected velocities of the components and accretion disk size.
The next generation of large sky area surveys, such as The Vera Rubin Observatory (LSST) and Euclid, are expected to find and follow-up thousands of lensed quasars from which such events could be identified and observed.
In this work we present a characterization and estimation of all HMEs that could potentially be observed, focusing on systems that could be identified by ground based telescopes.
From systems whose minimum image separation is at least 1 arcsec, and their second dimmest image is at least 21.5 magnitudes in the i-band ($\sim560$ in the southern or northern sky), we estimate $\sim60$ HMEs with amplitudes $>0.3$ [mag] in the r-band per year.
We find that on average, saddle images are approximately four times more likely to host events than minima, and $\sim10\%$ ($\sim50\%$) of events are caustic crossings for saddles (minima).
We also find that HMEs in saddle images can have amplitudes $\sim1-2$ [mag] larger than minima.}


   \maketitle
%

\section{Introduction}

It is well known that most massive galaxies host a supermassive black hole (SMBH) with mass around $\sim10^6-10^9\text{M}_\odot$ \citep{KormendyHo2013} at their core \citep{Richstone1998}.
A Quasar is believed to be a galaxy whose central SMBH is evolving as it grows in mass through accretion.
These SMBHs and their immediate surroundings ($\sim 0.01 - 10$ pc) are known as active galactic nuclei or AGN \citep[see][for a review]{Padovani2017}, and are believed to have a common angular structure.
The physical processes that take place in the AGN produce radiation pressure, winds, and jets that interact with the host galaxy and can thus affect its properties \citep[e.g.,][]{Hopkins2006,Fabian2012,Carniani2016}.
These processes, and the correlations that have been observed between the mass of the SMBH and the properties of the host galaxy \citep[see, e.g.,][]{Magorrian1998,Tremaine2002,Sun2019}, imply that their evolutions are tightly connected.
Hence, the models that describe the interactions between the AGN and its host galaxy throughout their joint evolution partly rely on our understanding of the accretion process in the AGN \citep[e.g.,][]{Heckman2014}.

The current consensus about the origin of the observed optical and UV power law continuum \citep[see Fig. 1.2 in][]{Harrison2014} in AGN is that it is due to friction-driven thermal emission from an accretion disk around the central SMBH.
Models that describe such disks, for typical parameters (e.g., accretion efficiency, Eddington luminosity, SMBH mass), suggest that they should have sizes on the order of $\sim1-10$ light days \citep[e.g.,][]{Mudd2018}.
Since the quasar luminosity function peaks at $z\sim2$ \citep{Palanque-Delabrouille2016}, the expected angular sizes are of the order of $\sim10^{-6}$ arcsec, which is well beyond our current observational resolution limits.
We expect these disks to exhibit a thermal slope, caused by increased friction in the inner regions due to denser faster-rotating material \citep[e.g., the standard thin disk model][]{ShakuraSunyaev1973}.
Because of this, if accretion disk sizes were measured using observations at shorter wavelengths, the sizes would be smaller than those measured at longer wavelengths \citep[see, e.g.,][]{Lira+2015}.
Therefore, the only way to measure accretion disk sizes for such distant objects is through indirect methods, such as reverberation mapping \citep{Fausnaugh2018} and microlensing in strongly lensed quasars \citep{Schmidt2010, Vernardos2024}.

Currently, the results of these two independent methods are in tension with the predictions of theoretical models.
In particular, they find accretion disk sizes that are larger and with shallower thermal slopes than predictions from a standard thin disk model \citep[e.g.,][]{Jimenez-Vicente2014,Fausnaugh2016,Bate2018,Fausnaugh2018,Jha2022}.
Both methods rely on modeling the observed AGN brightness variations over time.
Most AGN show intrinsic brightness variations due to inhomogeneities of the material being accreted \citep{Kawaguchi1998}.
The reverberation mapping technique relies on modeling the temporal correlation of observed brightness variations resulting from this intrinsic variability across different wavelengths and across different spectral features to probe the disk structure \citep[e.g.,][]{Cackett+2021}.
In the multiple images of strongly lensed quasars, we observe these intrinsic variations in every image shifted by a lensing-dependent time delay \citep{Refsdal1964}.
Coupled to this variation, an additional extrinsic component can be observed in all images due to lensing by the stars of the lensing galaxy.
These stars can further deflect the light from the quasar (on a $\sim10^{-6}$ arcsec scale) producing an additional strong lensing effect and thus (de)magnification of the individual image \citep[i.e., microlensing,][]{Kayser1986}.
Since the local distribution of stars in front of each lensed image is independent from the others, the brightness fluctuations due to microlensing are uncorrelated \citep{Paczynski1986}.
The amplitude of the brightness variation that the quasar image can experience depends on the size of the disk relative to the lensing scale of the microlens.
Because of this, and the fact that the size of the accretion disk depends on the observed wavelength, microlensing brightness variations have a chromatic signature, where smaller disks are more (de)magnified than larger ones \citep[e.g.,][]{Eigenbrod2008}.

Monitoring observations of strongly lensed quasars show microlensing variations with amplitudes of up to $\sim1$ mag on timescales from a few years up to decades.
The duration, amplitude and frequency of these variations depend on the many parameters that describe a lensed quasar system \citep[e.g., the relative velocities of the quasar, galaxy and stars, the stellar density in front of the lensed quasar image;][]{Neira2020}.
This means that these observations can in turn be used to infer many parameters of lensed quasar systems, including the projected dark matter fraction of the lensing galaxy \citep{Bate2011}, the peculiar transverse velocities of the lens and quasar \citep{PoindexterKochanek2010,Mediavilla2015}, and/or the quasar accretion disk size \citep{Morgan2010}.
However, the shorter (and thus steeper) variations, which we call high magnification events (HMEs), are most efficient for constraining the accretion disk structure \citep{Wyithe2000c}, especially when multi-wavelength data is available \citep{Anguita2008,Eigenbrod2008}.
Furthermore, it is of particular interest when a HME is due to the quasar crossing or passing close to a micro-caustic.
This is due to their infinite magnification and the overall high magnification around them.
These HMEs could potentially reveal inhomogeneities or special structures that the inner regions of the disk could have \citep[e.g.,][]{Lewis2004,Dexter2011,Yan2014,Best2022}, which in turn could help us have a better understanding of accretion physics.
\cite{Millon2022} show how during a HME these features can be probed.

Studies that rely on HMEs are hindered by two observational problems.
First, the necessary conditions for a quasar to become strongly lensed imply that these kinds of systems are rare, and thus only a few systems are known to date \citep[$\sim350$\footnote{This number is orders of magnitude lower than other astrophysical phenomena.} confirmed;][]{More+2016,Anguita+2018,Lemon2018,Lemon2022}.
Second, the long timescales of moderate to no microlensing variability before a HME occurs impose the necessity of monitoring these systems over periods of time on the order of decades \citep[e.g.,][]{Goicoechea2020}.
Although there are monitoring campaigns for strongly lensed quasars \citep[e.g.,][]{Millon2020, Munoz+2022}, HMEs rarely occur; coupled with the few systems they observe, this leaves HME studies confined to a minimum number of systems.
Most HMEs have been observed in the system Q 2237+0305 \citep{Huchra1985}, which is considered the ideal system to perform HME studies \citep[see, e.g.,][]{Wambsganss1990, Wyithe2000c, Anguita2008, Eigenbrod2008, Zimmer+2011, Odowd+2011, Goicoechea2020}.
This is due to the low redshift of its lensing galaxy ($z=0.04$), which enhances the projection of the transverse velocity of the microlenses relative to the quasar, while also decreasing the projected size of its quasar accretion disk relative to the lensing scale of the microlenses.
These two effects combined make microlensing events unusually short (from months to a year) and fairly more common (every $\sim4-5$ years).

With the forthcoming next generation of large area surveys, we expect to increase the number of identified lensed quasars to the order of thousands \citep{OM10}.
Furthermore, for the southern sky, the Vera Rubin Observatory Legacy Survey of Space and Time \citep[LSST][]{LSST} will provide multi-wavelength monitoring for many lensed quasars.
\cite{Neira2020} showed that a uniform observing strategy (i.e. the cadence at which different regions of the sky are re-visited) would be best to measure HMEs.
However, it is possible that the LSST will not follow a uniform cadence during the ten-year duration of the survey \citep{Bianco2022}, which could have a significant impact on the fraction of HMEs that it could observe.
This means that complementary observations to perform HME studies will likely be required.
Furthermore, for the science case of accurately probing the quasar accretion disk from light curves, even the best LSST strategy to observe a HME will likely not have a sufficiently high cadence to accurately constrain the desired parameters \citep[see, e.g., Fig. 2 in][]{Neira2020, Best2022}.
Here, triggering microlensing-tailored follow-up observations will also be required.
On the other hand, for the northern systems where a LSST-like survey is not available, selecting a subsample of systems with higher likelihood of hosting a HME may be a key step in performing microlensing studies.

To know what we could potentially expect in terms of observing HMEs, and thus prepare for the data that future surveys will provide, we present, analyze, and discuss the HME properties from a theoretically expected population of lensed quasars.
In Section 2 we present the models and assumptions required to generate simulated microlensing light curves.
Section 3 describes our definition of a HME along with a classification scheme.
In Section 4 we present the properties of the HMEs that have been identified in the simulated light curves.
We discuss in Section 5 the statistical differences between the different classifications of HMEs and predictions on what we should expect in terms of microlensing HMEs in lensed quasars.
We present our conclusions in Section 6.

\section{Methodology}
\label{sec:methodology}
The number of currently known lensed quasars is too small to use as a statistically significant population.
Therefore a simulated population of lensed quasar systems is required, by assuming a (lensing) galaxy and (background) quasar population \citep[see][]{OM10}.
Several assumptions are required in order to simulate the brightness variations due to microlensing.
In particular, a model of the mass distribution of the lensing galaxy (macromodel), the structure of the accretion disk of the quasar and the relative velocities of the different components are required.
This will result in a theoretically possible ensemble of unique systems with fixed parameters.
Furthermore, in order to capture each system HME variance, we need to simulate a large number of light curves
These parameters can then be fed to the tool presented in \cite{Neira2020}.
Even though this tool generates microlensing light curves and then samples them according to some specific LSST observing cadence, we will only consider the theoretical (i.e. non-sampled) light curves.
Here we describe the assumptions for each of the used models and their parameters that are required to use this tool.

\subsection{Simulated systems}
\label{sec:simulatedsystems}
The \cite{OM10} catalog (hereafter OM10) provided us with a theoretically expected population of lensed quasars.
Each lensed quasar system is composed of a lensing galaxy with an elliptical mass distribution and a background quasar. The distribution of their properties (i.e., redshifts, brightnesses, lens velocity dispersion) are consistent with what has been measured from observations.
The OM10 catalog provided us with most of the parameters needed for the models required by the tool presented in \citet{Neira2020} to generate the microlensing light curves.
Specifically, the parameters provided by the catalog and used in this work are the following:
\begin{itemize}
    \setlength{\itemindent}{1em}
    \item apparent magnitude in the i-band of the lensed image (m$_i$);
    \item location of the image relative to the lens ($x_i,y_i$);
    \item central velocity dispersion of the lens ($\upsilon_\text{disp}$);
    \item orientation and ellipticity of the lens ($\phi_e$, $e$);
    \item direction and magnitude of the external shear ($\boldsymbol{\gamma}_\text{ext}$);
    \item redshifts of the lens and source (z$_\text{l}$, z$_\text{s}$).
\end{itemize}

We selected systems that have a minimum image separation of 1 arcsec and whose second dimmest quasar image has an apparent magnitude lower than 21.5 in the i-band.
These constraints were set specifically to select systems that can be targeted with smaller (1 meter class) telescopes while ensuring that we always selected at least two images per system.
Thus, the selected systems could be studied without entirely relying on data from surveys such as the LSST, and instead also rely on data from observations from different telescopes.
We ended up with a subpopulation of $\sim$2800 lensed quasars out of the complete OM10 catalog, of which $\sim$2300 are doubles, $\sim$10 are naked cusps and $\sim$500 are quadruples.
We note however, that this number is an overestimate by a factor of 5 with respect to that expected in the southern or northern sky.
This choice was made to minimize the statistical noise.
Thus, in the following, any statistical property that is affected by this is scaled appropriately by a factor of 5 (i.e., as computed from the $\sim560$ systems expected in $\sim20\,000 \text{ deg}^2$).
As in \cite{OM10}, in this work we adopt a flat $\Lambda$CDM cosmology with $H_0 = 72.0$ km s$^{-1}$ Mpc$^{-1}$, $\Omega_m = 0.26$, and $\Omega_\Lambda = 0.74$.

\subsection{From the macromodel to magnification maps}
\label{sec:magmap}
The brightness variations due to microlensing can be studied by using magnification maps, which are generated through the inverse ray-shooting method \citep{Kayser1986}.
Because generating thousands of such maps is a computationally expensive task, we make use of the already generated maps of the GERLUMPH project \citep{GERLUMPH2014,GERLUMPH2015}.
The appropriate selection of a magnification map depends on the projected mass distribution of the lens at the location of the lensed image to be studied.
More specifically, magnification maps are defined by the values of the convergence ($\kappa$), shear ($\gamma$) and smooth matter fraction (s) at the specific location in the lens plane where the lensed images are.
These quantify the amount of matter, image distortion and amount of stars, respectively.
To obtain these values at the location of the quasar images in our sample and thus select the appropriate magnification map, we follow the same approach as in \cite{Foxley-Marrable2018,Vernardos2019}, whose main points we outline below.

Following the OM10 catalog definitions, we describe the mass distribution of all the lensing galaxies as a singular isothermal ellipsoid (SIE).
For a SIE, the convergence and shear at any point in the lens plane can be obtained through the following equation:
\begin{equation}
    \label{eq:kappa}
    \kappa(x,y) = \gamma(x, y) = \frac{1}{2} \frac{\theta_\text{E}\sqrt{q}}{\omega(x,y)} \text{.}
\end{equation}
Here $\theta_\text{E}$ and $q$ are the Einstein radius and axis ratio of the lens respectively, $\omega(x,y)$ is the elliptical radius and $x,y$ define the coordinates of the lens plane with the x-axis aligned with the major axis of the lens.
The Einstein radius of a lens whose mass distribution is described by a SIE potential is given by
\begin{equation}
    \theta_\text{E} = 4\pi \left(\frac{\upsilon_\text{disp}}{c} \right)^2 \frac{D_\text{LS}}{D\text{S}} \lambda(q) \text{,}
\end{equation}
where $\upsilon_\text{disp}$ is the velocity dispersion of the lensing galaxy, $c$ is the speed of light, $D_\text{LS}$ and $D_\text{S}$ are the angular diameter distances from the lens to the source and from the observer to the source respectively.
The parameter $\lambda(q)$ is a dynamical normalization value that arises from considering a two-dimensional lens potential from a three-dimensional shaped galaxy \citep[see equations 6, 7 and 8 in][]{Oguri2012}.
The total shear at any location in the image plane is the vector sum of the shear due to the lens potential ($\boldsymbol{\gamma}_\text{SIE}$), and an external component due to environmental effects ($\boldsymbol{\gamma}_\text{ext}$) \citep[see equations 3, 4 and 5 in][]{Vernardos2019}.
The magnitude of the external shear, $\gamma_\text{ext}$ is assumed to follow a log-normal distribution with mean 0.05 and dispersion 0.2, and its direction is assumed to be uniformly random \citep[see][]{OM10}.

The last component required to select a magnification map, the smooth matter fraction, can be computed as
\begin{equation}
    s(x,y) = 1 - \frac{\kappa_\star(x,y)}{\kappa(x,y)}
\end{equation}
where $\kappa_\star$ is the convergence due to compact matter (stars) only and can be computed from the light distribution and a mass-to-light ratio of the lens.
The following equation can be used:
\begin{equation}
    \label{eq:kappastar}
    \kappa_{\star,n} = A_n \exp{k_n \left( \frac{\omega(x,y)}{\sqrt{q}\theta_\text{eff}} \right)^{1/n}}.
\end{equation}
Here we have assumed a Sersic index $n=4$, $k_n=7.669$ \citep[see][]{Capaccioli1989} and $\theta_\text{eff}$ is the effective radius of the lens.
Lastly, $A_n$ is a normalization constant, where assumptions about the shape of the initial mass function (IMF) and the ratio of Einstein to effective radius are required \citep[see section 2.2 in][]{Vernardos2019}.

There are a total of 4 different set of assumptions, adopting either a Salpeter \citep{Salpeter1955} or Chabrier \citep{Chabrier2003} IMF, and a ratio of Einstein to effective radius derived from the Sloan Lens ACS \citep[SLACS]{Auger2009} or the CfA-Arizona Space Telescope Lens Survey \citep[CASTLES]{Oguri2014} lenses.
Analyzing and storing microlensing light curves of all of the systems for all different combinations of these assumptions would be computationally prohibited.
Thus, in this work we have limited ourselves to a Salpeter IMF and a ratio of Einstein to effective radius derived from the CASTLES lenses.
It is important to note that the role of the IMF here is purely related to the macromodel (i.e., the macro lens model), and not to the mass distribution of the microlenses.
Furthermore, the statistics of microlensing variability depend mainly on the mean mass of the microlenses \citep{Wyithe2001}.

\begin{figure}
     \includegraphics[width=\columnwidth]{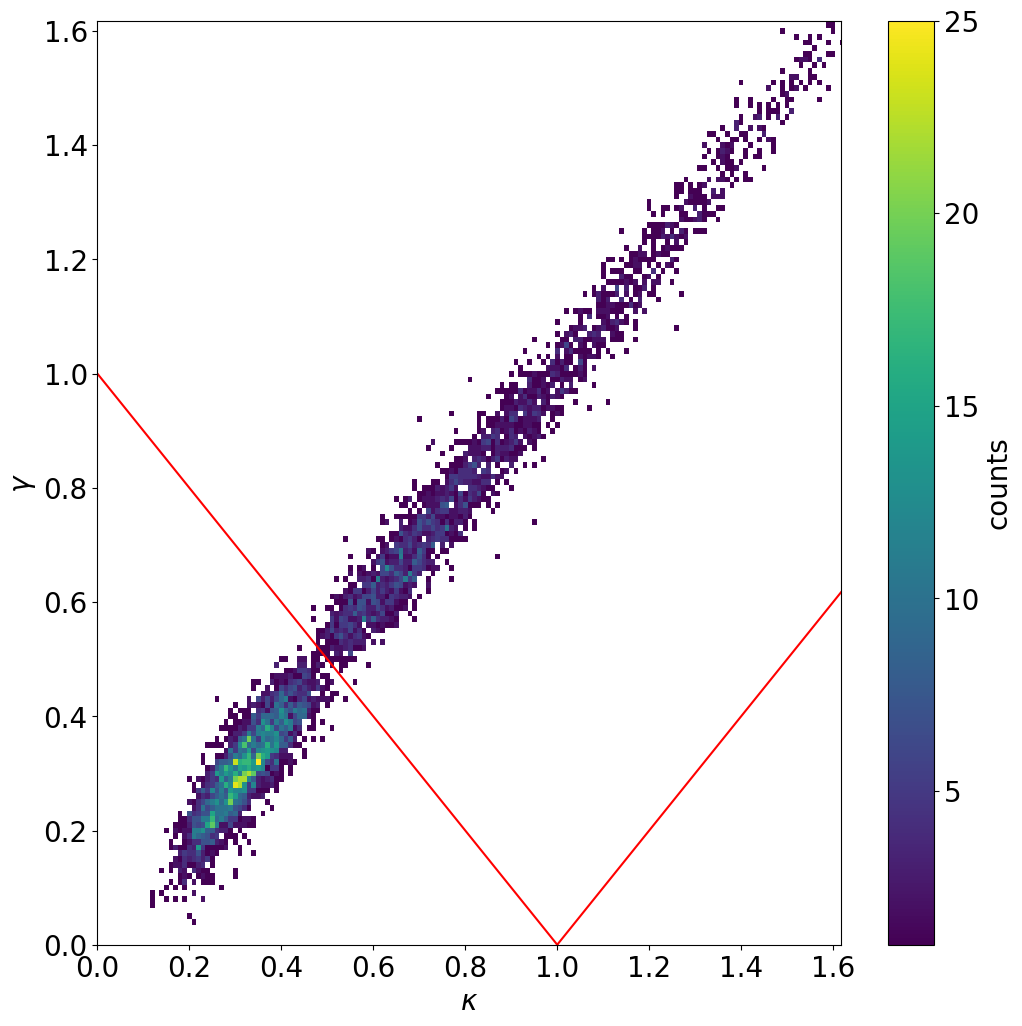}
     \caption{Two-dimensional histogram of $\kappa$ and $\gamma$ values computed for the selected lensed images from the OM10 catalog. The color in each bin represents the count number. The red solid line indicates where $\mu_\text{macro}\rightarrow\infty$ separating minimum (below) from saddle (above) images. We note that $\kappa\approx\gamma$ is due to describing the mass distribution of the lenses with a SIE + $\gamma_\text{ext}$. }
     \label{fig:macromodel}
\end{figure}

With this, we are able to compute the values of the convergence, shear and smooth matter fraction, and select the appropriate magnification maps from the GERLUMPH library for each lensed quasar image.
We show in Fig. \ref{fig:macromodel} the values of $\kappa$ and $\gamma$ for all images.
The magnification maps are of 10\,000 pixels side length, with a resolution of 0.025 $R_\text{E}/\text{pixel}$, where $R_\text{E}$ is the Einstein radius of the microlenses, given by
\begin{equation}
    \label{eq:remicro}
    R_\text{E} = \sqrt{\frac{D_\text{S}D_\text{LS}}{D_\text{L}} \frac{4G  M }{c^2}}\text{,}
\end{equation}
where $D_\text{L}$ is the angular diameter distance from the observer to lens, $G$ is the gravitational constant, $M=0.3[M\odot]$ is the mass of the microlenses (the approximate mean of a Salpeter IMF) and $c$ is the speed of light.

Additionally, for each magnification map we compute a corresponding caustic map by using a partially parallel implementation of the analytical method presented by \cite{Witt1990}.
Unlike the magnification map, this only contains the pixelated locations of the caustics (formally, points in the source plane where the magnification diverges).
Our use of this caustic map will be detailed in Section \ref{sec:mocklc}.

\subsection{Accretion disk}
\label{sec:accdisc}
In order to quantify microlensing in each lensed image, a physical model of the accretion disk is needed.
The detailed signatures of the microlensing brightness variations are tied to the specific model of the accretion disk \citep[see, e.g.,][]{Vernardos+2019,Best2022}.
However, because the goal of this paper is to study the properties of HMEs from a population of lensed quasars, and not the specific properties of different accretion disk models, we limit ourselves to a simple standard thin disk, ignoring any kind of substructure.
Thus, we have assumed an accretion disk size based on a standard thin disk model \citep{ShakuraSunyaev1973}
\begin{equation}
    \label{eq:size}
    R_\lambda = 9.7 \times 10^{15} \left(\frac{\lambda_\text{rest}}{\mu m}\right)^{4/3} \left(\frac{M_\text{BH}}{10^9M_\odot}\right)^{2/3} \left(\frac{f_E}{\eta}\right)^{1/3} [\text{cm}]\text{,}
\end{equation}
where $R_\lambda$ is the size of the disk at $\lambda_\text{rest}$, $M_\text{BH}$ is the mass of the black hole, $f_E$ is the Eddington ratio and $\eta$ is the accretion efficiency.
We adopted fixed typical values of $f_E=0.25$ and $\eta=0.15$ \citep[e.g.,][]{Blackburne2011}, while the black hole mass has been computed based on \cite{Macleod2010} as
\begin{equation}
    \text{log10}(M_\text{BH}) = 2 - 0.27 \times M_\text{i}\text{,}
\end{equation}
where $M_\text{i}$ is the absolute magnitude of the source on the i band \citep[K corrections from][]{Richards2006}.
Given a black hole mass estimate, we can compute the size of the accretion disk at any wavelength corresponding to the 6 photometric LSST bands - u, g, r, i, z and y ($\lambda_\text{eff}=$ 3654.9, 4800.3, 6222, 7540.6, 8682.1 and 9925 \AA{} respectively) -  through equation (\ref{eq:size}).
For simplicity, we match the sizes computed from equation (\ref{eq:size}) to the half light radii of Gaussian-shaped two-dimensional light distributions \citep[][show that the influence of the specific shape of the profile has only minor influence in the microlensing signal.]{Mortonson2005,Vernardos+2019}.

\subsection{Velocity model}

The microlensing brightness variations are due to the relative transverse movement of the quasar disk across the magnification map.
We must take in consideration the relative transverse velocities of the observer $\upsilon_\text{o}$, microlenses $\upsilon_\star$, source $\upsilon_\text{s}^\text{pec}$ and lens $\upsilon_\text{l}^\text{pec}$, projected onto the source plane \citep{Kayser1986}.
By combining these velocity components we can thus compute the final effective transverse velocity as the following vector sum:
\begin{equation}
    \label{eq:eff_v}
    \boldsymbol{\upsilon}_\text{e} = \frac{\boldsymbol{\upsilon}_\text{o}}{1+z_\text{l}}\frac{D_\text{LS}}{D_\text{L}} - \frac{\boldsymbol{\upsilon}_\star}{1+z_\text{l}}\frac{D_\text{S}}{D_\text{L}} + \boldsymbol{\upsilon}_\text{g}.
\end{equation}
Here we have combined the values of the peculiar velocities of the lens and source as $\boldsymbol{\upsilon}_\text{g}$.
We can do this, as we assume they follow a normal distribution with widths $\sigma_\text{l}^\text{pec}$ and $\sigma_\text{s}^\text{pec}$ for the lens and source respectively, with a uniformly random direction.
Thus $\boldsymbol{\upsilon}_\text{g}$ is drawn from normal distribution with width
\begin{equation}
    \label{eq:peculiar}
    \sigma_\text{g} = \left[ \left(\frac{\sigma_\text{l}^\text{pec}}{1+z_\text{l}} \frac{D_\text{S}}{D\text{L}}\right)^2 + \left(\frac{\sigma_\text{s}^\text{pec}}{1+z_\text{s}}\right)^2  \right]^{1/2}.
\end{equation}

The OM10 catalog does not constrain the values of $\sigma_\text{l}^\text{pec}$ and $\sigma_\text{s}^\text{pec}$, therefore we need to assume their values.
The dispersion of the peculiar velocity distributions can be modeled as
\begin{equation}
    \sigma^\text{pec}(z) = \frac{\sigma^\text{pec}(z=0)}{(1+z)^{1/2}} \frac{f(z)}{f(z=0)} \text{,}
\end{equation}
where $\sigma^\text{pec}(z=0)=235$ km/s \citep{Kochanek2004}, $f(z)$ is the growth rate function which \cite{Lahav1991} show that can be approximated by
\begin{equation}
    f(z) \approx \Omega_\text{m}(z)^{0.6} \text{,}
\end{equation}
and
\begin{equation}
    f(z=0) \approx \Omega_\text{m}(z=0)^{0.6} + \frac{\Omega_\Lambda}{70}\left(1+\frac{\Omega_\text{m}(z=0)}{2}\right) \text{,}
\end{equation}
where $\Omega_\text{m}$ and $\Omega_\Lambda$ are the matter and dark energy density parameter, respectively.

The transverse velocity of the observer, $\upsilon_\text{o}$, can be precisely defined as measured from the cosmic microwave background (CMB) velocity dipole \citep{Kogut1993}. It can be computed as
\begin{equation}
    \boldsymbol{\upsilon}_\text{o} = \boldsymbol{\upsilon}_\text{CMB} - (\boldsymbol{\upsilon}_\text{CMB} \cdot \hat{z})\hat{z},
\end{equation}
where $\upsilon_\text{CMB}$ is the CMB velocity and $\hat{z}$ is the vector along the line of sight.
Thus, the magnitude and direction of $\boldsymbol{\upsilon}_\text{o}$ depends on the location in the sky of the lens system.
Because the OM10 catalog does not constrain the position in the sky of the systems, we need to assign values of right ascension and declination for each system, which will constrain $\upsilon_\text{CMB}$ (and thus $\upsilon_0$).
We therefore place the systems uniformly random in the sky.
We note that \cite{Neira2020} show that depending on the location on the sky, the observer velocity can make for up to $\sim$50 percent of the contribution of the total effective transverse velocity of the system.

The final component, that is the random movement of the microlenses, would formally require the use of non-static magnification maps.
However, \cite{Wyithe2000a} show that this can be statistically approximated by a bulk velocity for all the microlenses as:
\begin{equation}
    \upsilon_\star = \sqrt{2}\epsilon\sigma_\text{disp} \text{,}
\end{equation}
where $\sigma_\text{disp}$ is the velocity dispersion at the center of the lens and $\epsilon$ is an efficiency factor that depends on the macromodel (here we assume $\epsilon=1$).
This velocity component due to the movement of stars only adds in magnitude to the vector sum of the previous components (hence why it is called a bulk velocity).
Therefore it is treated as a velocity that is constant in magnitude with an identical direction to the vector sum of the other random velocity components.

\subsection{Mock light curves}
\label{sec:mocklc}
The parameters of the models presented above are used as input in the microlensing light curve generator tool \citep{Neira2020}.
For each lensed image we simulate sets of 100\,000 10-year-long light curves in each of the 6 photometric LSST bands.
Additionally, using the corresponding caustic map (see the end of section \ref{sec:magmap}), for each set of light curves, we generate a corresponding caustic curve (as well as a touching a caustic curve), which has a value of 1 when the center (any region within the half light radius) of the disk is touching a caustic and a value of 0 elsewhere.
Thus, for each of the 100\,000 randomly generated tracks in each lensed image, we will have a final set of thirteen curves: six light curves, six touching curves and one caustic curve.
We note that, unlike in \cite{Neira2020}, here we do not take in consideration any observing strategy, and only analyze the continuous theoretical curves.

\section{Microlensing events}
\label{sec:events}

\subsection{Defining the events}
Typically, HMEs are qualitatively described as peaks of brightness in a light curve.
HME properties such as duration and amplitude have only been quantitatively measured by calculating the full width at half maximum of these peaks \citep{WambsganssKundic1995}.
The frequency of these peaks can depend on the size of the disk relative to the Einstein radius and/or the effective transverse velocity.
These can be such that peaks occur over long periods of time; therefore, long monitorings would be required to probe them.
This means that, for a fixed probed time, even if the brightness significantly changes, without a brightness peak an event would not be identified.
Motivated by this, we instead define a HME as any brightening (magnification) or dimming (de-magnificaton) in the simulated light curve that has an amplitude larger than an arbitrarily defined threshold.
We begin by finding all the extrema points in the light curve.\footnote{We also consider the first and last point in time of the light curve as extrema.}
We then compute the change in brightness between every possible combination of pairs of extrema points, for example, the first and second, the first and third, second and third.
Any pair of these points can potentially correspond to a HME, if and only if the following criteria are fulfilled:
\begin{itemize}
    \item The amplitude of their brightness variation is larger than the arbitrarily defined magnitude threshold.
    \item Within the time interval defined by this pair of points there are no extra pairs of extrema points (including pairs with any of the two points defining the time frame) with a brightness variation of the same sign and with a larger amplitude.\footnote{In this way we find the largest in amplitude events.}
    \item Within the time interval defined by this pair of points there are no extra pairs of extrema points (including pairs with the two points defining the time frame) with a brightness variation of the opposite sign and with an amplitude of at least the magnitude threshold.\footnote{If this were the case, it would mean that there are at least two events. As an example, let us consider four points in time $t_a < t_b < t_c < t_d$ that satisfy the magnitude threshold condition, where the pair of points $(a, d)$ have a positive amplitude. If the amplitude of the pair $(b, c)$ is a negative, then the event that starts at $a$ ends at $b$, thus there would be three events defined by the pairs $(a, b)$, $(b, c)$ and $(c, d)$.}
\end{itemize}
These conditions imply that: i) any brightness variation with an amplitude smaller than the arbitrarily defined magnitude threshold is treated as noise, and ii) HMEs start immediately after the previous event ends in an alternating manner (i.e. a magnification is followed by a de-magnification and vice versa).

Because we maximize the amplitude of the HME, we are likely to end up with HMEs where for a significant amount of time there are no major brightness variations.
To circumvent this, we redefine the starting (ending) times of magnification (de-magnification) events after the initial identification.
The magnification (de-magnification) events are effectively shortened such that they start (end) when $5\%$ ($95\%$) of their maximum flux variation is reached.
Because we define a HME to be either a magnification or de-magnification, and redefine the starting (ending) times of magnification (de-magnification) events, we note that our definition can yield two contiguous events that meet at a peak of brightness (magnification followed by de-magnification), or two events that are disconnected at their tails (de-magnification followed by magnification).\footnote{An implementation of the algorithm used to find these events, as well as some examples, can be found in \url{https://github.com/fcneirad/event_finder}.}

\subsection{Characterizing the events}
\begin{figure}
     \includegraphics[width=\columnwidth]{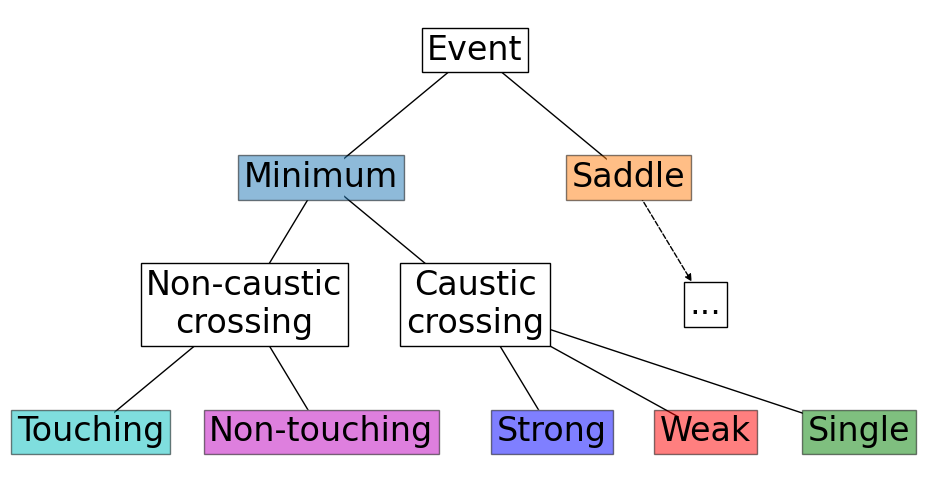}
     \caption{Classification scheme of a microlensing HME.}
     \label{fig:scheme}
\end{figure}

Because we are simulating light curves from a population of lensed quasars with a wide range of properties, we should expect a large variety in the properties of the HMEs that we will identify.
To help us characterize HMEs, we have created a classification scheme (see Fig. \ref{fig:scheme}) based on two main attributes.
The first one is the parity of the macro-image to which the HME belongs.
This is motivated by the fact that the parity of the macro-image is tightly correlated with the micro magnification that a source can experience.
This has been explored by \cite{Schechter2002} and further expanded into a significantly larger macromodel parameter space by \cite{VernardosFlukeb2014}.
They show that a notable difference between saddle and minimum images is that in the former, the source can be significantly more de-magnified in correlation with the smooth matter fraction.
We should thus expect microlensing brightness variations over time in saddle images to have larger amplitudes than in minimum images.

The second main property for HME classification is whether or not the event is produced by a caustic crossing.
We define such events as when the center of the disk crosses at least one caustic during the HME.
This classification is independent of the image parity classification and done by using the caustic curve defined in section \ref{sec:mocklc}.
The caustic crossing classification is motivated by the fact that caustics have very high magnification, and thus have the potential to reveal the specific structure of the accretion disk.
Furthermore, the average magnification in the close vicinity of a caustic is different when measured inside the closed high magnification region defined by the caustic than outside.

Generally, the brightening and dimming produced by a caustic crossing event shows an asymmetric profile.
In this scenario, given our definition of HMEs, for each caustic crossing two contiguous caustic crossing HMEs would be identified, where one would have a significant larger amplitude than the other.\footnote{Note that this is not always the case: for source sizes that are $\lesssim R_\text{E}$ (which is true for most cases here), as it crosses through a cusp, the event will look roughly symmetric. However, since cusps have significantly smaller cross section than folds, it is more likely that a caustic crossing occurs through a fold.}
With this in mind, we further classify caustic crossing events into three categories.
When two contiguous events are identified, they will be characterized as a ``strong'' and ``weak'' event.
The event with the largest amplitude is classified as a strong caustic crossing event.
For events of this kind, the finite size of the source and the difference in magnification between the inside and outside of the closed caustic region play a significant role.
We can describe these events as those where we can resolve the caustic, i.e., the asymmetric feature is apparent, and that the disk crossing the caustic typically happens during the strong caustic crossing event.

The third category of caustic crossing events is used if only one event is identified (i.e., without another contiguous one).
In this case, the event will be classified as a single caustic crossing event.\footnote{The word single is used to refer to the fact that the event does not have a companion, not that the event is due to the disk crossing a single caustic.}
There are two reasons for such events to happen, either the amplitude of a contiguous weak caustic crossing event is smaller than the minimum threshold (thus it remains unidentified\footnote{If this weak event would have been identified, the single caustic crossing event would have been classified as a strong caustic crossing event, hence the unidentified weak event.}), or simply because there is no contiguous event.
In the former scenario the event can be easily recovered by lowering the minimum threshold to identify HME.
For the latter, non-existing contiguous events can be explained in two ways: i) in some systems, the size of the disk, in combination with its effective velocity is such that, the brightness variations occur over very long periods of time, or ii) the events occur at the very end or start of the light curve.
In this scenario, more observing time would be needed in order to observe a peak of brightness.
Therefore, and analogously to the strong and weak caustic crossings interpretation, due to the lack of such peak (and thus no asymmetric-like feature), we can describe single caustic crossing events as ones where the caustics are not resolved.

Finally, we also classify non-caustic crossing events as touching and non-touching a caustic, where we define touching as any region within the half light radius of the disk touching a caustic during a HME.

In summary, we have defined four main classifications for HME, three subclassifications for caustic crossing events and two subclassifications for non-caustic crossing events, yielding a total of 10 classifications.
Our classification scheme is shown in Fig. \ref{fig:scheme}.
Additionally, in Appendix \ref{app:A} we show light curve examples, where we have identified their HMEs and highlighted their corresponding (non-)caustic crossing classification.
We do not highlight the parity classification as this pertains to the macromodel parameters, and not the light curve itself.

\section{Results}
\label{sec:results}
We identify HMEs in the 10-year long simulated light curves in the r-band, which is typically most efficient for observing quasars \citep[e.g.,][]{Eulaers2013}.
We apply our classification scheme to these events using a magnitude threshold of 0.3 mag, which is reasonably large to be detected by most telescopes.
We have computed the number, duration, and amplitude of HMEs from an ensemble of systems in all the simulated lensed quasar images.
Here, we present the HME properties differences between their classifications.

\subsection{Image parity}
\label{sec:imgparity}
\begin{figure}
 \includegraphics[width=\columnwidth]{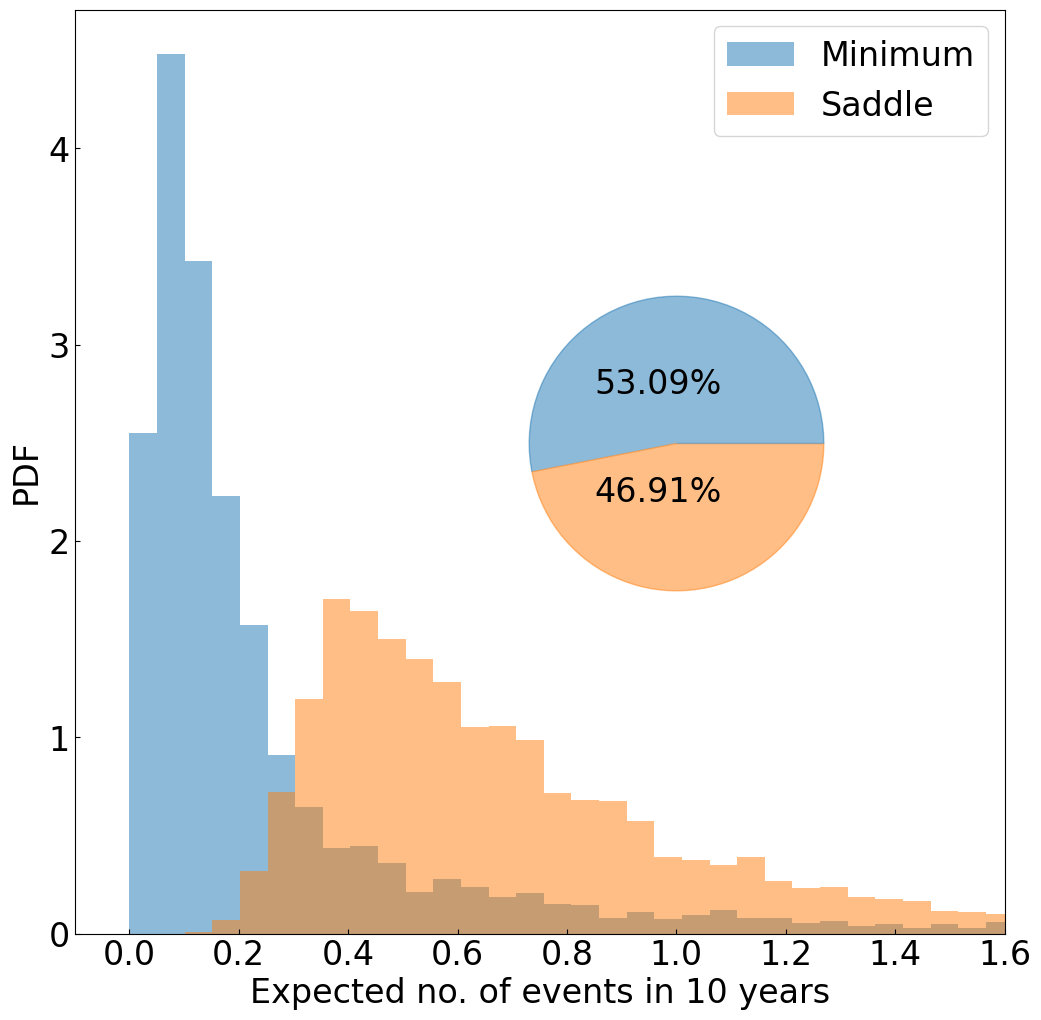}
 \caption{Expected number of events in 10 years per image parity in all 6190 lensed images. The distribution of saddle and minimum images peak at $\sim$0.4 and $\sim$0.1 respectively. Because the area of each histogram is independently normalized, the pie chart represents the total area fraction of each histogram, i.e. $\sim53\%$ of the images are minimum.}
 \label{fig:nevents}
\end{figure}

In Fig. \ref{fig:nevents}, we show the distribution of the expected number of events (i.e., the mean number of events per curve in each image) for either image parity.
These follow a log-normal distribution defined as
\begin{equation}
    f(n, \mu, \sigma) = \frac{1}{\sigma n \sqrt{2\pi}} \exp{\left(-\frac{\log^2(n/\mu)}{2\sigma^2}\right)},
\end{equation}
with $n$ the expected number of events, $\mu = 0.16$, $\sigma = 1.06$ for minimum and $\mu = 0.63$, $\sigma = 0.59$ for saddle images.

\begin{figure}
 \includegraphics[width=\columnwidth]{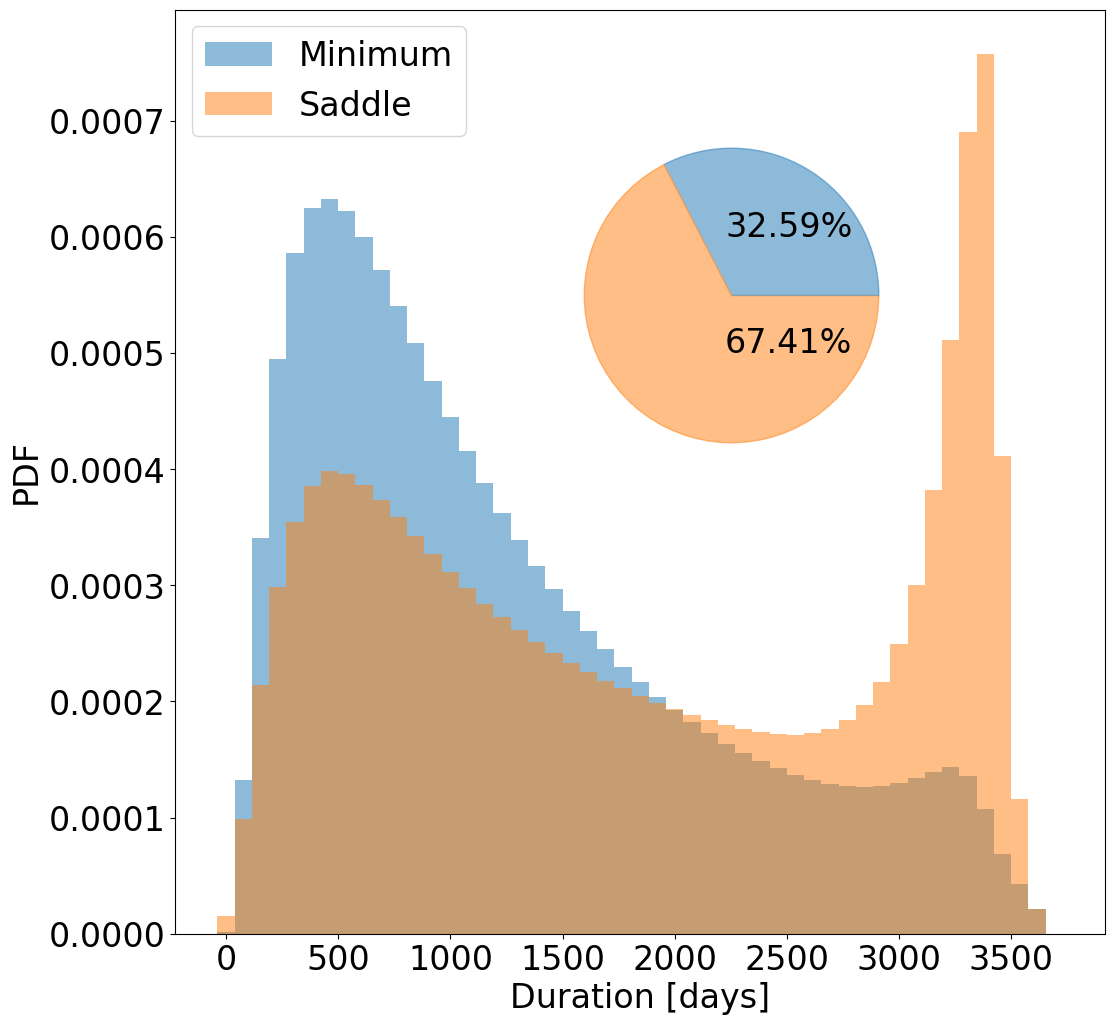}
 \caption{Duration of all HMEs per image parity. We note that the peak at $\sim3300$ [days] corresponds to events that last longer than the 10 years of the light curves. As per Fig. \ref{fig:nevents}, the pie chart represents the total area fraction of each histogram, i.e. $\sim32$\% of the events belong to a minimum image.}
 \label{fig:length_all}
\end{figure}

\begin{figure}
 \includegraphics[width=\columnwidth]{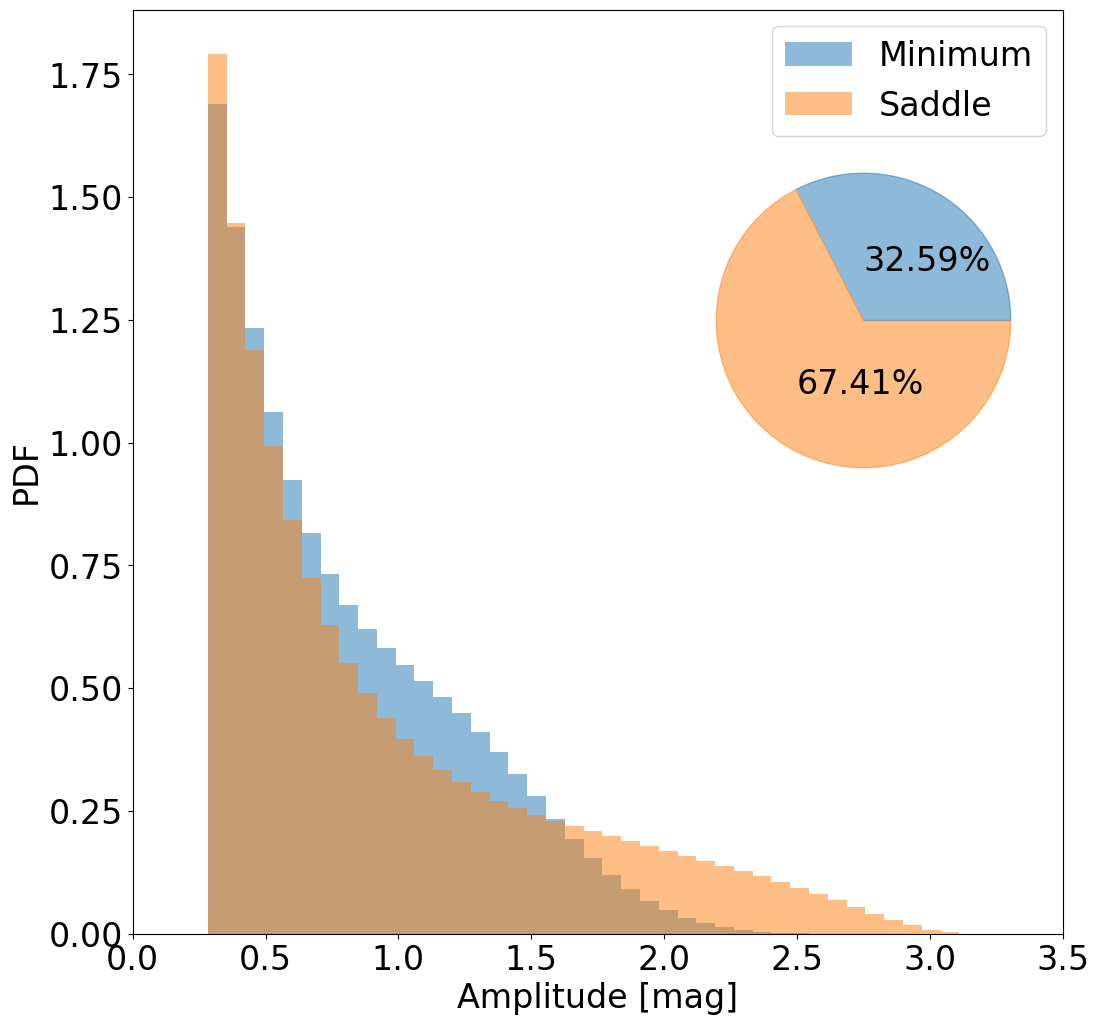}
 \caption{Amplitude of all HMEs per image parity. We note that the larger amplitudes in saddle images are due to their magnification maps having deeper de-magnification regions.}
 \label{fig:strength_all}
\end{figure}

We show the distribution of the duration of all HMEs in Fig. \ref{fig:length_all} which shows two peaks, at $\sim500$ and $\sim3300$ [days].
The latter one corresponds to HMEs that last longer than the 10-year sampled period of the evaluated light curve.
We note that this peak at $\sim3300$ [days] is not at $\sim3650$ [days] (the 10 years sampled) due to our shortening of the events explained in Section \ref{sec:events}.
A notable difference between the two image parities, is the fact that the peak that corresponds to this kind of event is larger in saddle images than in minimum ones.
We also note that Figs. \ref{fig:length_all} and \ref{fig:strength_all} show that minimum images host $\sim33\%$ of all HMEs, but from Fig. \ref{fig:nevents} (and according to the log-normal fit) they only host $\sim25\%$.
This discrepancy is due to the fact that in Figs. \ref{fig:length_all} and \ref{fig:strength_all} show the distribution of all HMEs, while Fig. \ref{fig:nevents} shows the distribution of the expected number of events in each quasar image (i.e., the mean number of events per quasar image.)

We show the amplitude of all HMEs in Fig. \ref{fig:strength_all}.
Naturally, events with a smaller amplitude are more common.
Thus, the amplitude distribution of HMEs shows a peak at 0.3 magnitudes, the magnitude threshold used to identify the events.


\subsection{Caustic and non-caustic crossings}
\label{sec:results_cc}

\begin{figure}
 \includegraphics[width=\columnwidth]{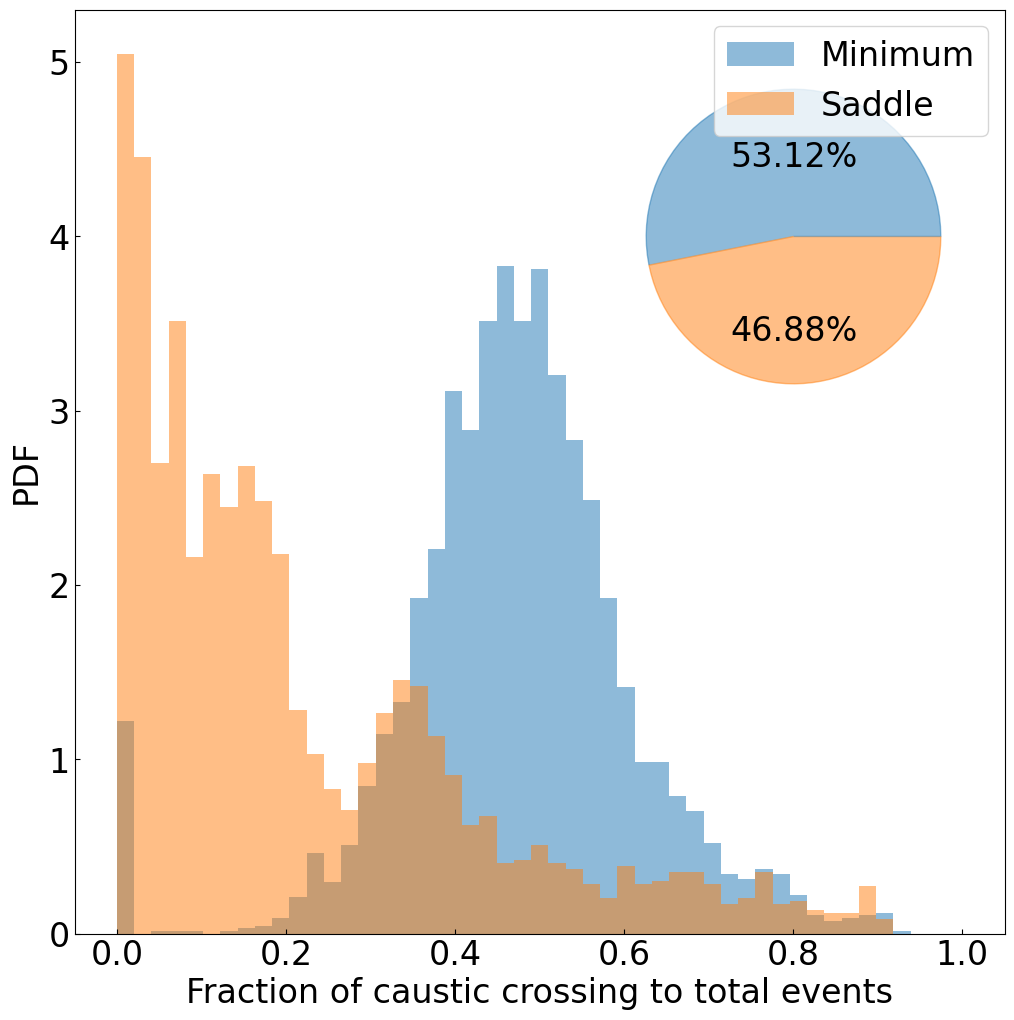}
 \caption{Fraction of events that are caustic crossings per image parity. Fractions in saddle and minimum images peak at $\sim0.1$ and $\sim0.5$ respectively. As per Fig. \ref{fig:nevents}, the pie chart represents the total area fraction of each histogram, i.e., $\sim53$\% of the images are minimum. We note that the fractions shown in Fig. \ref{fig:prob_cc} and \ref{fig:nevents} are not exactly the same. This is due to the fact that in a very few number of images there are no HMEs identified, thus their fraction of caustic crossings remains undefined.}
 \label{fig:prob_cc}
\end{figure}

The fraction of HMEs that are caustic crossings is shown in Fig. \ref{fig:prob_cc}.
Here, we find that an event is significantly more likely to be a caustic crossing when it occurs in a minimum image.

\begin{figure}
 \includegraphics[width=\columnwidth]{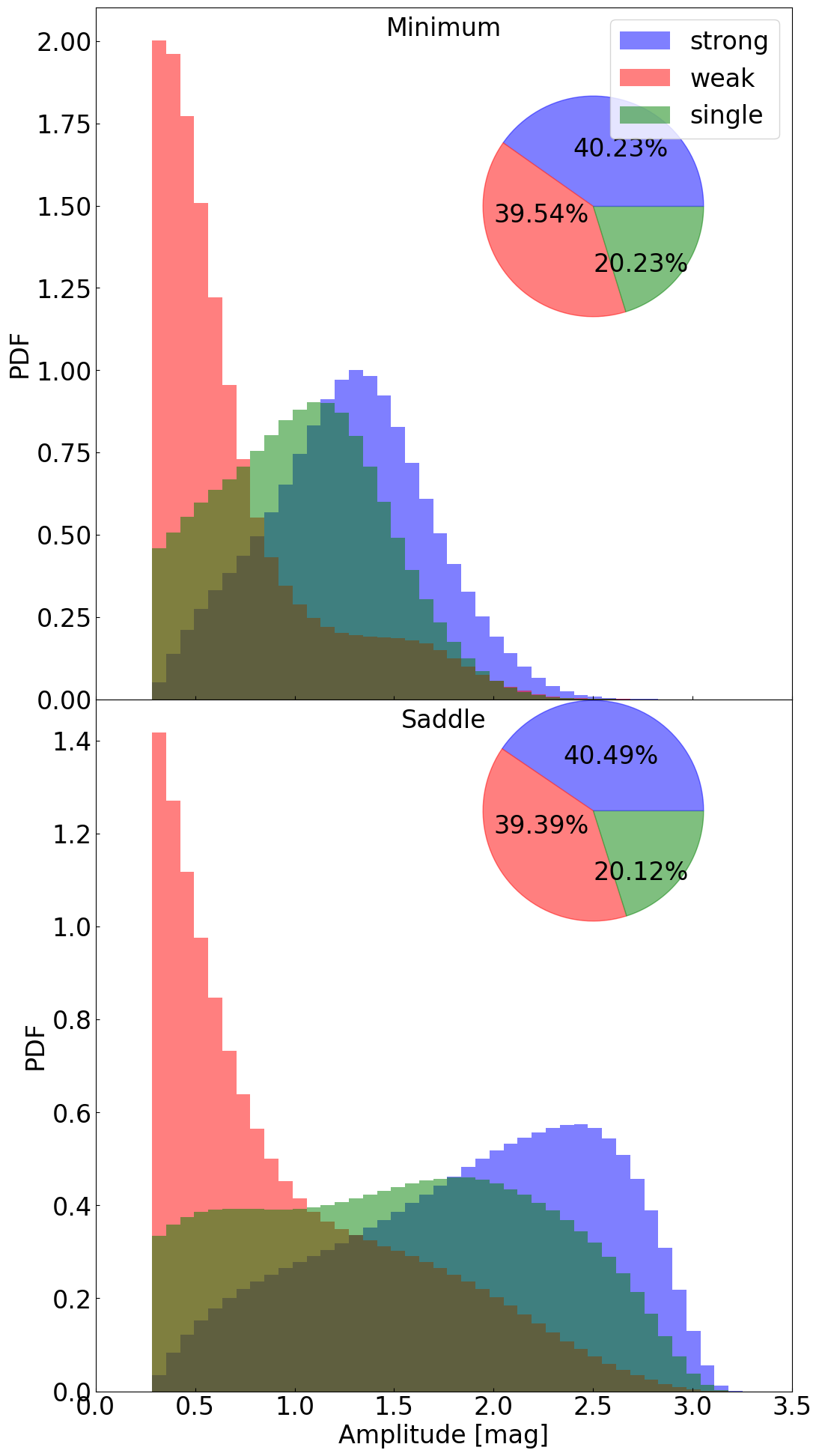}
 \caption{Amplitude of all caustic crossing events for all the minimum (top) and saddle (bottom) images. The colors match the scheme classification of caustic crossing events shown in Fig. \ref{fig:scheme}.}
 \label{fig:strength_cc}
\end{figure}

\begin{figure}
 \includegraphics[width=\columnwidth]{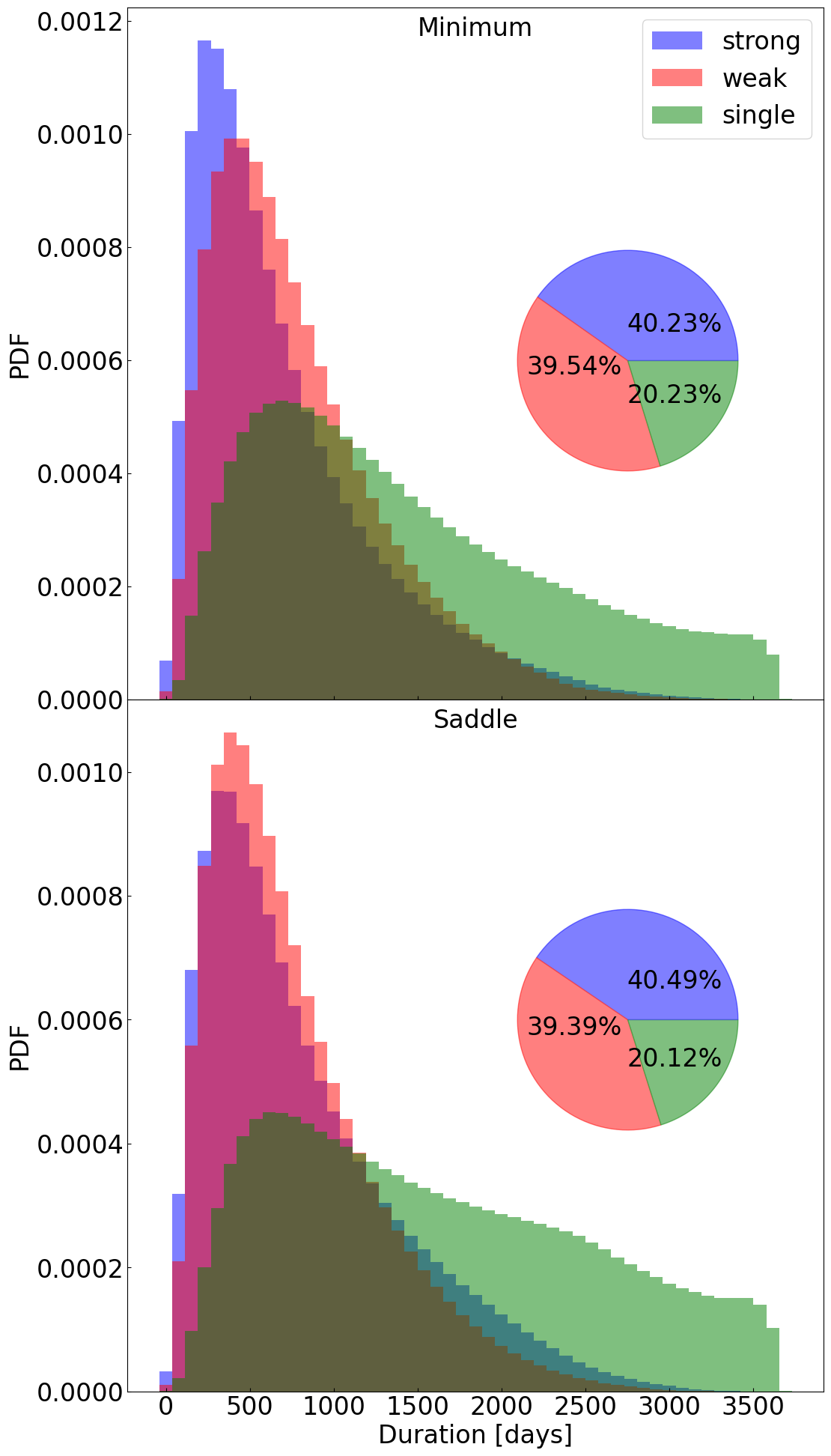}
 \caption{Duration of all caustic crossing events for all the minimum (top) and saddle (bottom) images. The colors match the scheme classification of caustic crossing events shown in Fig. \ref{fig:scheme}.}
 \label{fig:length_cc}
\end{figure}

We show in Fig. \ref{fig:strength_cc} the amplitude for the three kinds of caustic crossings that we have defined, where we have color coded strong, weak and single events as blue, red and green, respectively (consistent with the colors in Fig. \ref{fig:scheme}).
Naturally, strong and weak events have a larger and smaller amplitude respectively, while for single events, the amplitude distribution spans over the combined range of strong and weak.
Furthermore, the differences between image parity seen in Fig. \ref{fig:strength_all} become much more pronounced, particularly in the strong caustic crossings, where saddle images can be $\sim$1 to $\sim$1.5 magnitudes larger than in minimum images.
As for their duration, we show the distributions in Fig. \ref{fig:length_cc}.
Interestingly, and in contrast to the differences seen in the amplitudes of caustic crossings, the differences in the duration of caustic crossings between image parity are minimal.

\begin{figure}
 \includegraphics[width=\columnwidth]{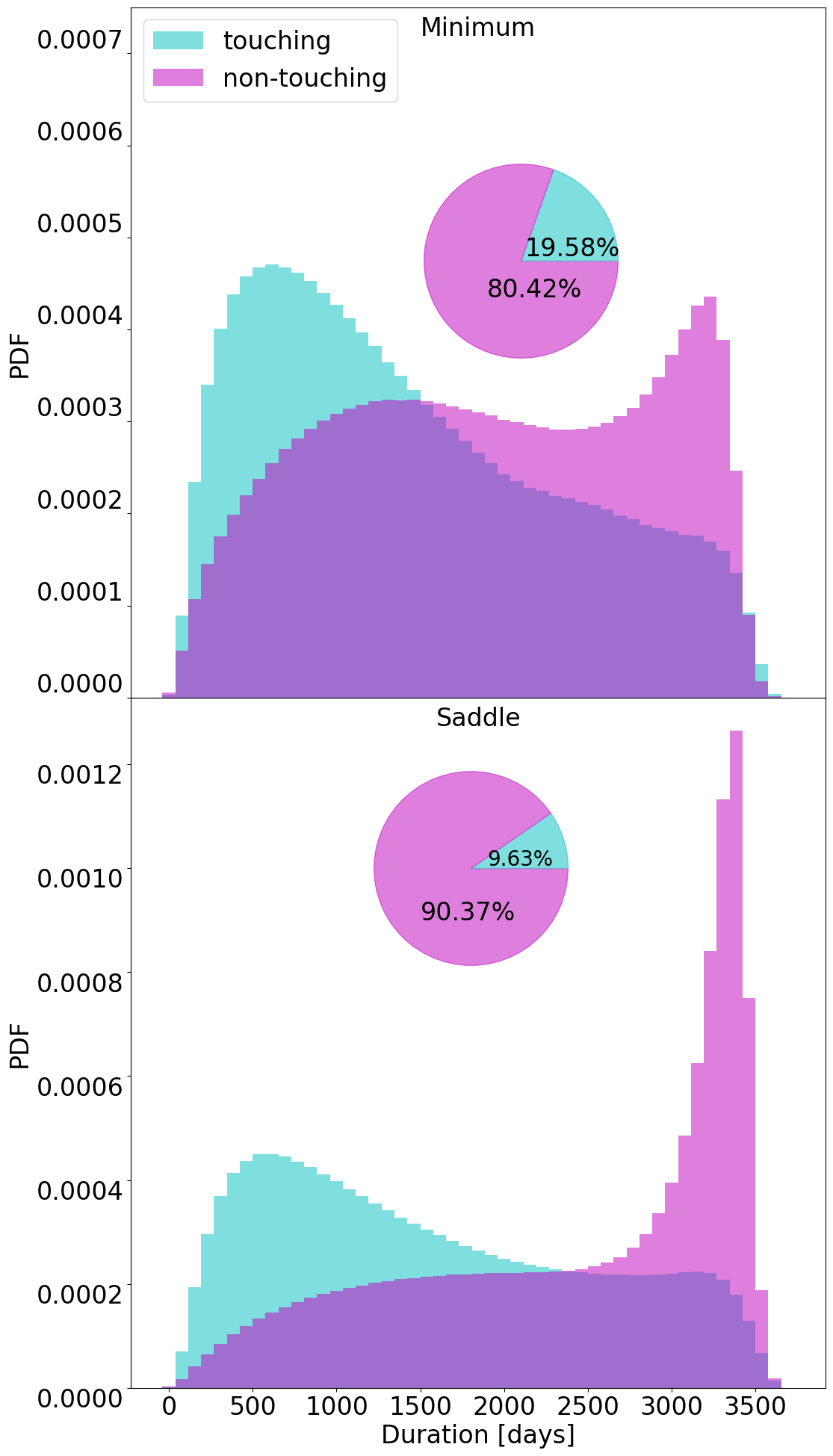}
 \caption{Duration of all non-caustic crossing events for all minimum (top) and saddle (bottom) images. The colors match the scheme classification of non-caustic crossing events shown in Fig. \ref{fig:scheme}.}
 \label{fig:length_ncc}
\end{figure}

\begin{figure}
 \includegraphics[width=\columnwidth]{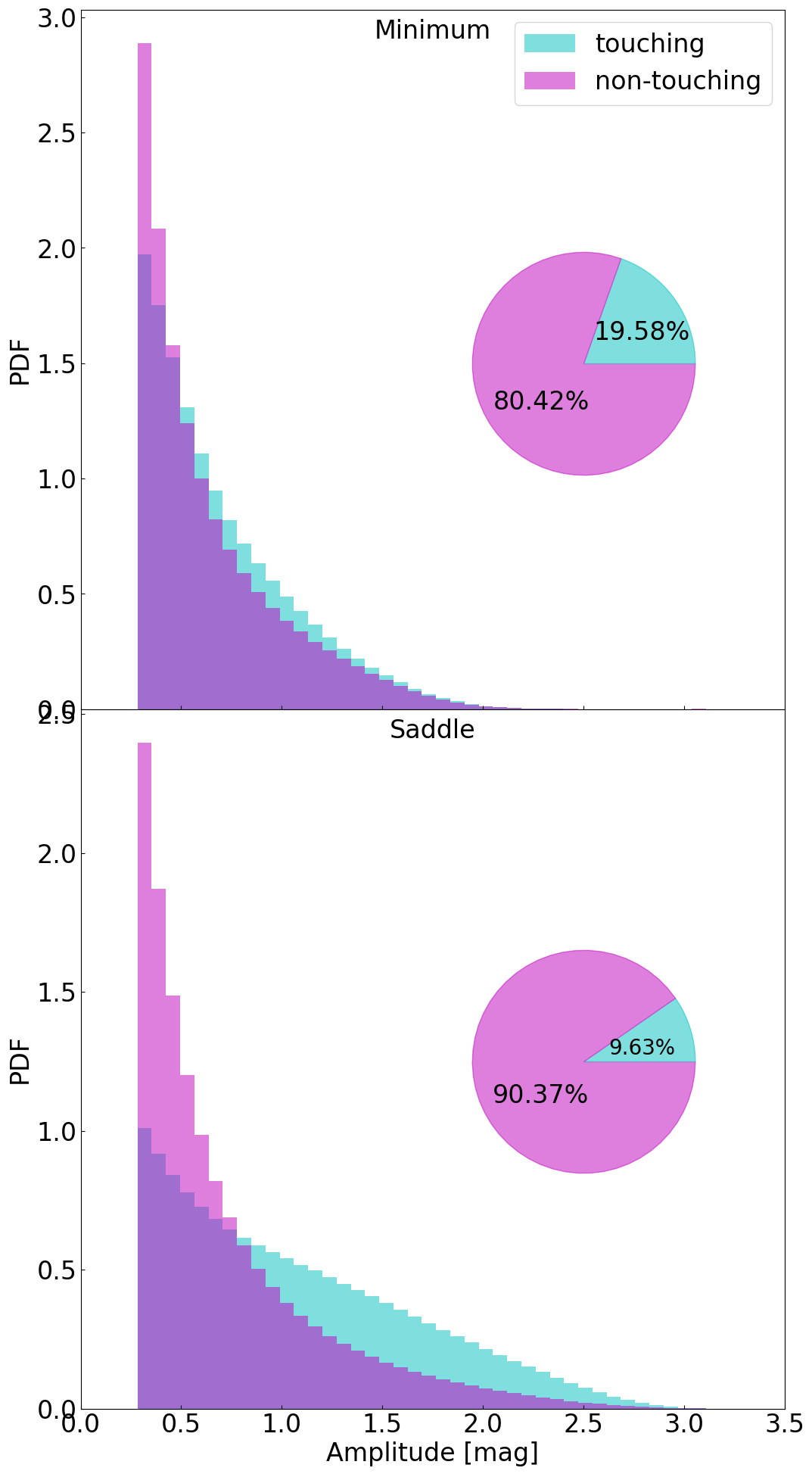}
 \caption{Amplitude of all non-caustic crossing events for all minimum (top) and saddle (bottom) images. The colors match the scheme classification of non-caustic crossing events shown in Fig. \ref{fig:scheme}.}
 \label{fig:strength_ncc}
\end{figure}

We also show the duration and amplitude of non-caustic crossing events, after applying their respective subclassification (touching and non-touching, non-caustic crossings) in Fig. \ref{fig:length_ncc} and \ref{fig:strength_ncc} respectively.
Here we see how for both image parities, the duration of a HME where the disk touches a caustic can be significantly shorter than when it does not.
Furthermore, we see that for saddle images there are significantly more longer than shorter non-touching non-caustic crossing HMEs.
For the amplitude of these two kinds of HMEs, we see how for minimum images there are minimal differences, while for saddle images the touching events are more likely to have a larger amplitude.

\section{Discussion}

The forthcoming big surveys are expected to increase the number of known strongly lensed quasars by at least an order of magnitude \citep{OM10}.
\cite{Neira2020} broadly predicts $\sim300$ HMEs with amplitudes >1 [mag] per year to be potentially observed by the LSST.
There are however, a few caveats to consider for their estimation in comparison to the study presented here.
First, they consider fainter systems, up to 24 magnitude in the i-band (single epoch detection for LSST), whilst here we consider only systems up to 21.5 magnitude.
Thus, we only take into account a fraction of the systems they consider.
And second, their rough estimation is made under the assumption that all lensed images of the systems that are to be discovered would have microlensing properties of image A of the lensed quasar HE 0230-2130.
By analyzing the events on a population of lensed systems, rather than on a single lensed image, in this work we are able to make more realistic estimates about what we should expect of quasar microlensing HMEs in the near future.

\subsection{Image parity and caustic crossings}

The broader range of micro magnification that saddle images exhibit \citep[compared to minimum images; e.g.,][]{VernardosFlukeb2014}, in particular the deeper de-magnification regions, has three main consequences in the brightness variations of microlensing light curves.
First, these de-magnification regions imply that, for a random trajectory of a source, we should expect more brightness variations, and thus on average more HMEs, as shown in Fig. \ref{fig:nevents}.
This is because, the source no longer needs to move through high magnification regions for an event to occur.
Instead it can move from a deep de-magnification to a shallower de-magnification to have a significant brightness variation.
Secondly, we should also expect the amplitude of such variations to be larger , as shown in Fig. \ref{fig:strength_all}.
Lastly, these de-magnification regions also imply that it is not necessary for the source to actually cross a caustic for a HME to occur, as shown in Fig. \ref{fig:prob_cc}.
Therefore, we should expect saddle images to have on average more and larger amplitude HMEs, but also that many of these events will not be caustic crossings.

In Fig. \ref{fig:length_all} we can see a notable difference in the duration of HMEs between saddle and minimum images.
We observe two peaks in the distributions of both image parities, where the peak at $\sim3300$ [days] is due to events that would last longer than the 10 years of the simulated light curve.
The fact that this peak is much larger for saddle than for minimum images can be explained by two main reasons:
i) the larger expected number of HMEs, and ii) the lower fraction of caustic crossing events in saddle images.
This implies that non-caustic crossing events are more common in saddle images.
Contrary to caustic crossing events, where it is more common to identify two contiguous events that meet at the brightness peak, non-caustic crossing events typically do not have a contiguous one, and therefore a peak of brightness would not be observed or identified.
This can be seen in Fig. \ref{fig:length_ncc}, where the non-caustic crossing events are significantly more likely to span over the 10 years of the simulated light curve.
We note that if the source travels close to a cusp, two events can be identified, where a brightness peak in the light curve comes from the high magnification region that lies close to the caustic.
We can see this again in Fig. \ref{fig:length_ncc}, where the shorter events are more likely to be due to a region of the disk touching a caustic.

The resulting distributions of the duration and amplitude of all HMEs can be explained by how a closed (micro) caustic can be.
For the following, it is important to make the distinction between the caustic curve, and its immediate surroundings (i.e. inside and outside the curve).
The nature of the curve itself, in terms of its magnification, remains mostly invariant along the parameter space of the macromodel of the images (except for the shear, which effectively increases the high magnification area defined by a closed caustic).
This means that the parity of the image does not play a significant role.
Hence the differences in the duration of HMEs between saddle and minimum images are minimal (Fig. \ref{fig:length_cc}).
On the other hand, the effects in the surroundings of a caustic, namely the (micro) (de)magnification, do indeed depend on the parity of the image, where saddle images exhibit deeper de-magnification regions.
Consequently, caustic crossings in saddle images can have, in average, from 1 to 1.5 magnitudes larger amplitudes than in minimum images.
Furthermore, the difference in magnification between the caustic itself and its surroundings in the outside and inside of the area described by it, is typically larger for the former.
Because of this, strong and weak caustic crossing events, would correspond to when the disk is outside and inside of the caustic respectively (see, e.g., Fig. \ref{fig:example_strong_weak}).
This in not always the case, for regions of high caustic density, where the difference in amplitude between strong and weak caustic crossing events may not be as large compared to isolated caustics.
This is reflected in the overlap in the amplitude distribution between strong and weak events (see Fig. \ref{fig:strength_cc}).

\subsection{HMEs in the future}

From the oversampled set of simulated lensed quasar systems in the OM10 catalog ($\sim2800$), from which we have generated microlensing light curves, a fifth ($\sim560$) corresponds to what we expect to find, given our observational constraints explained in Section \ref{sec:simulatedsystems}.
This amounts to about $\sim1250$ lensed quasar images split between doubles and quads.
Based on this number, through a bootstrap analysis, we can make a realistic estimate of the expected number of HMEs that could potentially be observed.
We randomly select a fifth of the systems and compute their average number of HMEs in 10 years.
We repeat this process until the mean number of HMEs converges, and we calculate the error due to the statistical noise of the OM10 catalog.
In addition to this error, we also take into account the variance due to the microlensing light curve parameter space (i.e., different regions of the magnification map, the sampled effective transverse velocity), and sum them in quadrature.
We estimate that in either the northern or southern sky (20\,000 sq. deg.) $61.5_{-8.9}^{+7.9}$ HMEs with a minimum amplitude of 0.3 [mag] in the r-band will occur each year, where the errors correspond to the $68\%$ confidence interval.

\begin{figure}
 \includegraphics[width=\columnwidth]{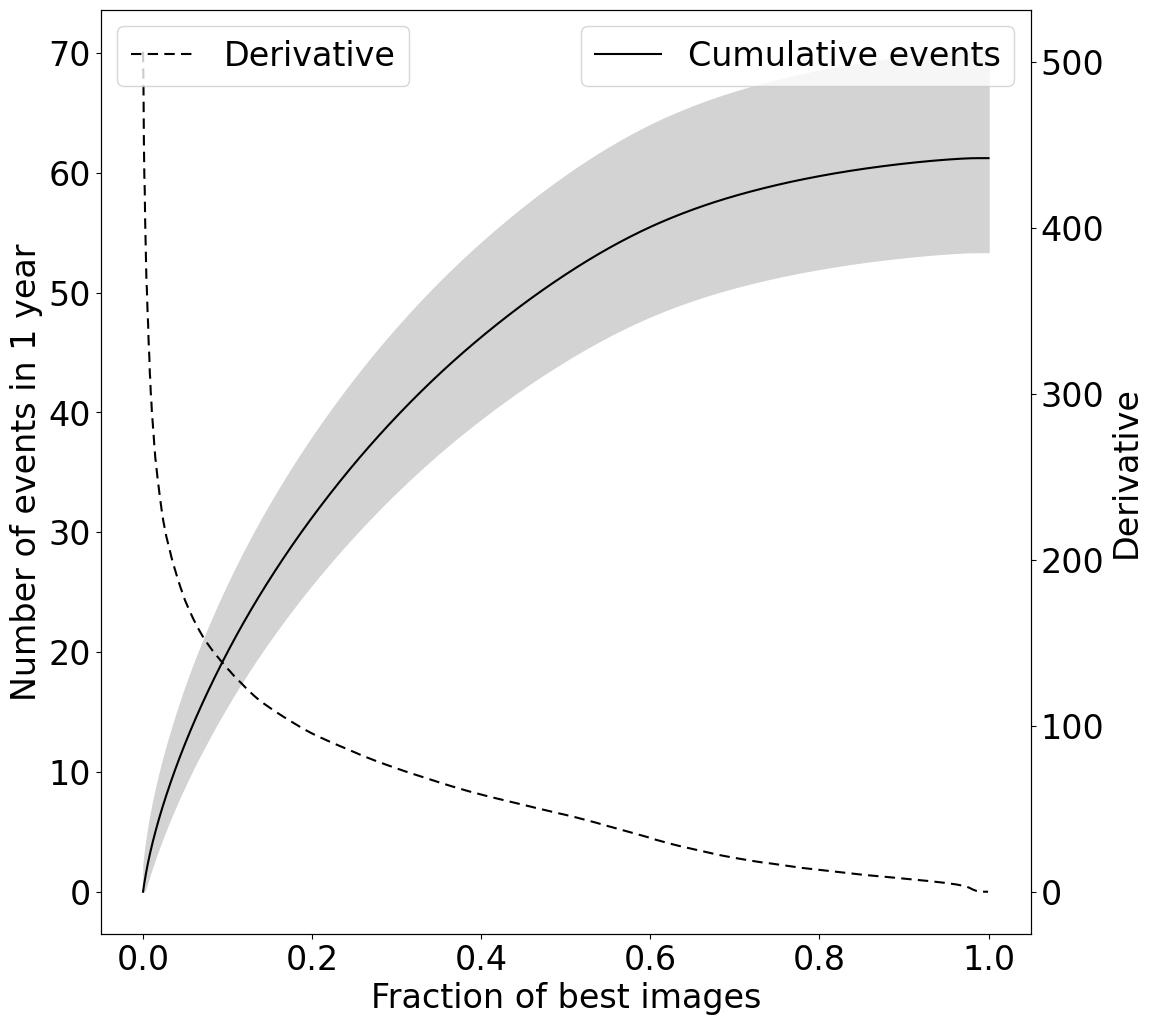}
 \caption{Expected number of events with an amplitude >0.3 [mag] in the r band as a function of the fraction the highest ranked (larger expected number of events) over the total number of images. The derivative is plotted in a dashed line. The shaded area corresponds to the $68\%$ confidence interval. The $\sim20\%$ of the highest ranked images have $\sim50\%$ of the number of HMEs.}
 \label{fig:best}
\end{figure}

The large variety of the properties of HMEs (i.e., expected number, duration) in each system, implies that selecting sub-populations of quasar images that are optimal to study will depend on the science goal (e.g., studying the source or lens).
In this regard, we chose the expected number of events as a metric to determine the suitability of a system to be used in microlensing studies.
With this in mind, we have ranked the quasar images by this metric, where the best images have a large expected number of HMEs.
In Fig. \ref{fig:best} we show the total number of events in 1 year as a function of the fraction of the highest ranked over the total number of images.
By applying an elbow-like criterion (i.e., when observing more quasar images does not significantly add more observed events) on its derivative we can estimate that by monitoring the best $\sim20\%$ images, $\sim50\%$ of the total number of events expected can be potentially observed.

\begin{figure}
 \includegraphics[width=\columnwidth]{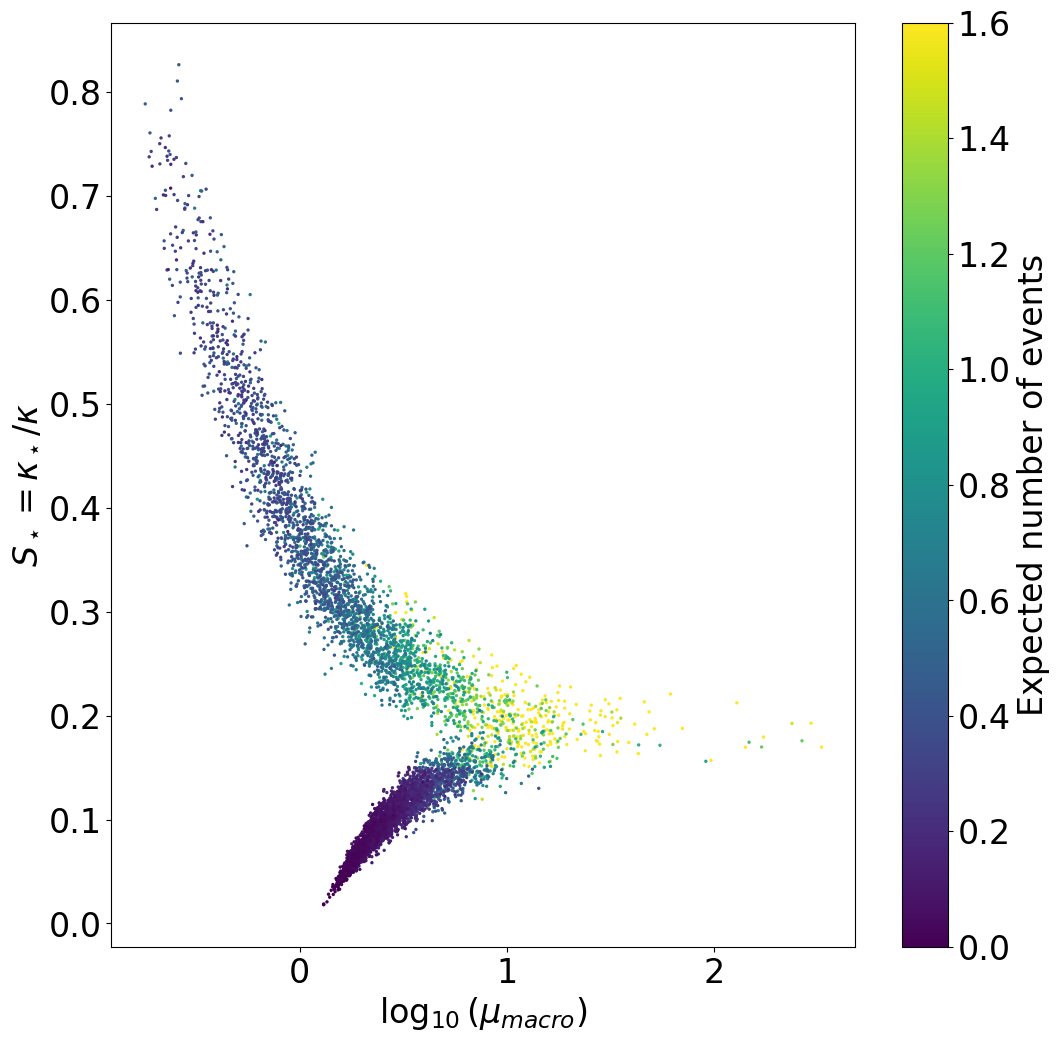}
 \caption{Expected number of HMEs in 10 years of all the lensed images as a function of the macro magnification $\mu_\text{macro}$ and compact matter fraction $S_\star$. Yellow (purple) indicates a higher (lower) expected number of HMEs. The expected number of events is strongly correlated with $\mu_\text{macro}$.}
 \label{fig:scores}
\end{figure}

We show in Fig. \ref{fig:scores} the expected number of events as a function of the macro magnification ($\mu_\text{macro}$) and compact matter fraction ($s_\star = \kappa_\star/\kappa = 1- s$).
We find that the expected number of events of a quasar image is correlated with its macro magnification.
One would also expect this number to find a correlation with $S_\star$, but this is weakened by the different macro-magnifications and extended sources \citep[see][for details]{Schechter2002}.
The best quasar images lie closer to the $\mu_\text{macro} \to \infty$ region, in agreement with the results presented by \cite{Weisenbach2021}.

Finally, HMEs also depend on other parameters (e.g. the source and lens redshift, the quasar magnitude, velocity parameters, etc.).
However, their dependency is not present, which can be explained by two main reasons.
First, their dependency is degenerate.
For example, decreasing the lens redshift and increasing the intrinsic quasar brightness would yield a faster but larger accretion disk.
While faster velocities imply more HMEs, a larger disk size implies less HMEs due to the disk having to travel longer distances in order to cross the caustics.
And second, the simulation is the result of an ensemble of these parameters.

\subsection{Caveats}

The results presented here depend on the models and parameters that we have adopted.
The degenerate nature of these parameters makes quantifying their effect on our results a non-trivial task without re-running the simulation.
Instead, here we qualitatively describe the caveats of our assumptions.

\subsubsection{Mock catalog}
The OM10 catalog adopts physical parameters derived from observations that were available at the time.
We note that a new catalog of gravitationally lensed quasars has been presented in \cite{Yue2022}.
This new catalog is built by adopting updated treatments on the quasar luminosity function and galaxy-velocity dispersion.
Their catalog predicts about $\sim30$ percent more systems than the OM10 catalog for quasars with redshift around $\sim2.2$.
However, \cite{Lemon2022} show that the number of lensed quasars even in this new catalog is under-predicted.
Nevertheless, since we show that the expected number of HME mostly depend on the macro magnification of the quasar images, we should expect that our results do not change qualitatively.

\subsubsection{Model assumptions}

Among our assumptions described in Section \ref{sec:methodology}, the ratio of Einstein to effective radius, together with the shape of the IMF, will only affect the smooth matter fraction (and thus affecting the magnification maps selected).
The former can be assumed from lenses from the CASTLES or SLACS surveys, while the latter as a Chabrier or Salpeter.\footnote{This is because these assumptions are derived from available observational fits, see section 2.2 in \cite{Vernardos2019}.}
Because the smooth matter fraction is anti-correlated with the caustic density of magnification maps, we should expect that assumptions that yield larger (smaller) fractions result in a smaller (larger) average number of caustic crossing events.
The differences in the derived smooth matter fractions between the different combinations of these assumptions can be seen in fig. 4 of \cite{Vernardos2019}.
However, because we find that the expected number of all HMEs is mostly correlated with the macro magnification of the quasar image, we should expect that these assumptions do not drastically change our results.
Finally, our accretion disk model has a direct impact on the duration and expected number of HMEs.
In this work we have adopted a thin disk model, which is simple enough and yet physically motivated.
Nevertheless, the size of the disk is the primary parameter that influences the microlensing variations \citep{Mortonson2005}.
Other accretion disk models would yield a greater number of faster events with larger amplitudes if these models yielded smaller sizes in comparison with the model used here.
Conversely, an accretion disk model that yields larger sizes would result in fewer slower events with smaller amplitudes.
Another assumption is that we have taken all accretion disks to be face-on.
In reality, most disks would have an inclination, producing a geometric effect which effectively reduces the source size \cite{PoindexterKochanek2010}, and consequently producing shorter and larger amplitude events.

We have qualitatively explained how our results are expected to change given different assumptions of our models.
However, the dependency of the properties of microlensing events presented here with physical parameters (e.g., redshift, disk size) is lost.
This is due to the degeneracy between them and the fact that we analyze multiple systems altogether which have different parameters.
Ultimately, ignoring the physically and empirically motivated assumptions, these event properties depend on the accretion disk size, effective transverse velocity and macromodel ($\kappa$, $\gamma$ and $s$) of the quasar image.

\subsubsection{Observational bias}
A final caveat for our estimations is due to a Malmquist-like bias explained in \cite{Baldwin2021}.
On average, the associated (macro) caustic area of quadruply imaged quasars is biased against those with smaller area ones, as it is directly proportional to the probability of being quadruply imaged.
While on the other hand, and because the area of the caustic is inversely proportional to the (macro) magnification of the quasar images, the depth of a survey produces a classical Malmquist bias against fainter (and thus against larger caustic area) systems.
The overall effect results in a smaller bias against small caustic areas, and therefore higher magnified systems are expected to dominate the known population.
Thus, and because we find that the macro magnification is strongly correlated to the expected number of HMEs, this implies that our prediction of events is an underestimation.

\begin{table}
\caption{Expected total number of events in one year for each combination of magnitude threshold and band.}
\def\arraystretch{1.2}

\centering

\begin{tabular}{c|c c c}

\hline
\multirow{3}{*}{Bands}             & \multicolumn{3}{c}{Threshold}\\ \cmidrule{2-4} 
  & $\Delta\text{mag}>0.3$ & $\Delta\text{mag}>0.5$ & $\Delta\text{mag}>1.0$ \\
\hline
\multirow{2}{*}{u}             & $68.8_{-9.4}^{+8.4}$                    & $51.0_{-8.2}^{+7.2}$                    & $29.2_{-6.5}^{+5.4}$         \\
                               & $36.9_{-7.2}^{+6.0}$                    & $28.9_{-6.5}^{+5.4}$                    & $17.6_{-5.3}^{+4.2}$         \\[1ex]
\multirow{2}{*}{g}             & $65.2_{-9.2}^{+8.1}$                    & $47.5_{-8.0}^{+6.9}$                    & $27.0_{-6.3}^{+5.2}$         \\
                               & $34.1_{-6.9}^{+5.8}$                    & $26.4_{-6.2}^{+5.1}$                    & $16.1_{-5.1}^{+4.0}$         \\[1ex]
\multirow{2}{*}{r}             & $61.5_{-8.9}^{+7.9}$                    & $44.4_{-7.7}^{+6.7}$                    & $24.9_{-6.1}^{+5.0}$         \\
                               & $31.5_{-6.7}^{+5.6}$                    & $24.2_{-6.0}^{+4.9}$                    & $14.9_{-4.9}^{+3.8}$         \\[1ex]
\multirow{2}{*}{i}             & $58.9_{-8.8}^{+7.9}$                    & $42.1_{-7.5}^{+6.5}$                    & $23.2_{-5.9}^{+4.8}$         \\
                               & $29.6_{-6.5}^{+5.4}$                    & $22.7_{-5.8}^{+4.7}$                    & $13.9_{-4.8}^{+3.7}$         \\[1ex]
\multirow{2}{*}{z}             & $57.0_{-8.6}^{+7.6}$                    & $40.5_{-7.4}^{+6.4}$                    & $21.8_{-5.7}^{+4.6}$         \\
                               & $28.4_{-6.4}^{+5.3}$                    & $21.7_{-5.7}^{+4.6}$                    & $13.3_{-4.7}^{+3.6}$         \\[1ex]
\multirow{2}{*}{y}             & $55.2_{-8.5}^{+7.4}$                    & $39.0_{-7.3}^{+6.2}$                    & $20.4_{-5.6}^{+4.5}$         \\
                               & $27.2_{-6.3}^{+5.2}$                    & $20.9_{-5.6}^{+4.5}$                    & $12.6_{-4.6}^{+3.5}$         \\
\hline
\end{tabular}

\tablefoot{For each combination, two numbers are given: events for all (top) and best 20 percent (bottom) lensed images are shown. The errors correspond to the $68\%$ confidence interval. It is important to note that the events quoted in redder bands are always identified in bluer ones, but not vice-versa.}
\label{tab:best}
\end{table}

\subsection{Additional bands and thresholds}
Up to this point, we have looked at HMEs that have been identified through a magnitude threshold of $0.3$ [mag] in the r band.
However, it is worth to explore other magnitude thresholds and bands.
Due to the large number of figures that would be required to show these results, we provide an online appendix, where the reader can statistically reproduce the plots shown here for different combinations of bands and thresholds (see the data availability statement).

In total we have analyzed the results from three thresholds: 0.3, 0.5 and 1.0 [mag], and the 6 photometric bands of the LSST: u, g, r, i, z and y, amounting to a total of 18 combinations of threshold and band.
We show the expected number of events in 1 year for each of these combinations in Table \ref{tab:best}.
Naturally, increasing the minimum threshold in magnitude to identify HME results in a decrease of the expected number of events.
However, and also expectedly, the fraction of events that are caustic crossings increases by $\sim20$ and $\sim40 \%$ for 0.5 and 1.0 threshold respectively.
Furthermore, for these larger thresholds the number of single caustic crossings decreases, in particular, the events that last at least the 10 years of the simulated light curve are no longer identified.
This is because these long events correspond to small brightness variations that result mostly from slow effective transverse velocities and/or large accretion disk sizes.

\section{Summary}

In this paper we have simulated $\sim 4$ billion light curves from a theoretically expected population of lensed quasars.
We have defined and identified HMEs in the resulting light curves and predict on the order of tens to a hundred HMEs each year could be potentially observed in the northern or southern sky.
The results presented here have allowed us to quantify the expected number and characteristics (e.g., amplitude, duration) of HMEs between minimum and saddle images.

We have made all the results available online to enable exploring the HMEs properties for a specific quasar image or subpopulation of the simulation (see the data availability statement).
Within the scope of upcoming large area monitoring surveys, this will certainly help understand what we should expect in the near future, and will allow us to make the necessary preparations for each specific science case.
One such example would trigger HME follow-up observations from the LSST.
Another example could focus on the northern sky, where there are no surveys like the LSST.
Comparing the expected properties of HME between specific systems could help prioritize the limited available observing time.

\section*{Data availability}
\subsection*{Other bands and thresholds}
In the body of the paper we presented Figures for the properties of HMEs in the r-band and threshold of 0.3.
We provide an online appendix at \url{https://github.com/fcneirad/OM10-HME}.
Here, the reader can reproduce the plots from Fig. \ref{fig:nevents} through \ref{fig:strength_ncc} for any combination of threshold and band.

\begin{acknowledgements}
      This work has been supported by ANID-FONDECYT Regular Project 1240105, ANID Millennium Science Initiative AIM23-0001, ANID BASAL project FB210003 and the Swiss National Science Foundation (SNSF). This research was made possible by the generosity of Eric and Wendy Schmidt by recommendation of the Schmidt Futures program.
\end{acknowledgements}

%
%
\bibliographystyle{aa}
\bibliography{ulens-om10}

\appendix
\onecolumn
\section{HME classification examples}
\label{app:A}
\begin{figure*}[h]
     \includegraphics[width=1\columnwidth]{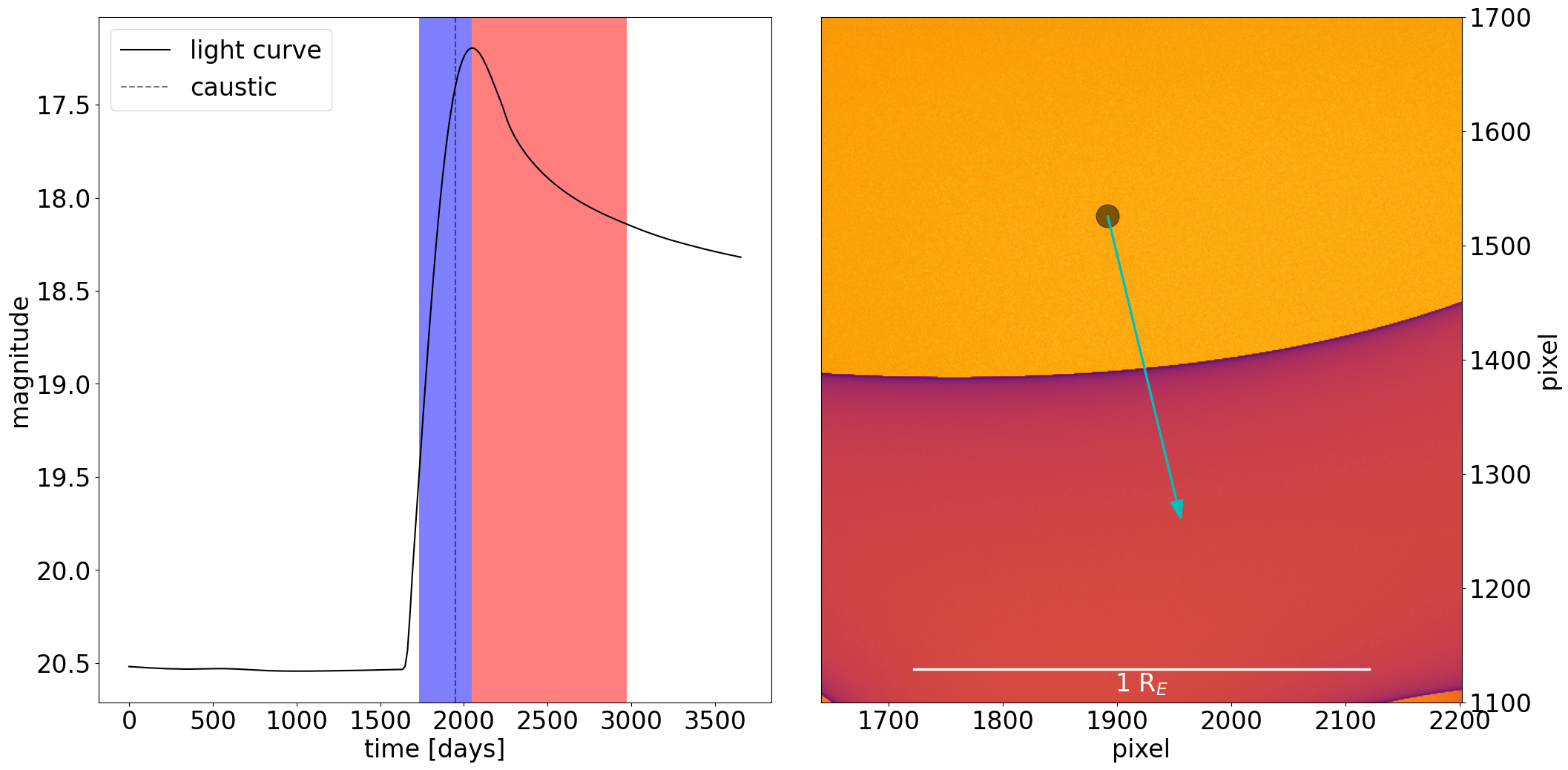}
     \caption{Left: Example of a simulated microlensing light curve. The vertical dashed lines indicate when the center of the disk crosses a caustic. The color shaded sections indicate the duration of a HME, where the color indicates its corresponding classification according to Fig. \ref{fig:scheme}.
     Right: Corresponding track of the disk (cyan arrow) on a magnification map. The darker (lighter) areas of the magnification map are high (low) magnification regions, while the shaded circle indicates the size of the accretion disk.
     We note the two identified events are contiguous and due to crossing a caustic, thus they are classified as strong and weak caustic crossings.}
     \label{fig:example_strong_weak}
\end{figure*}

\begin{figure*}
     \includegraphics[width=1\columnwidth]{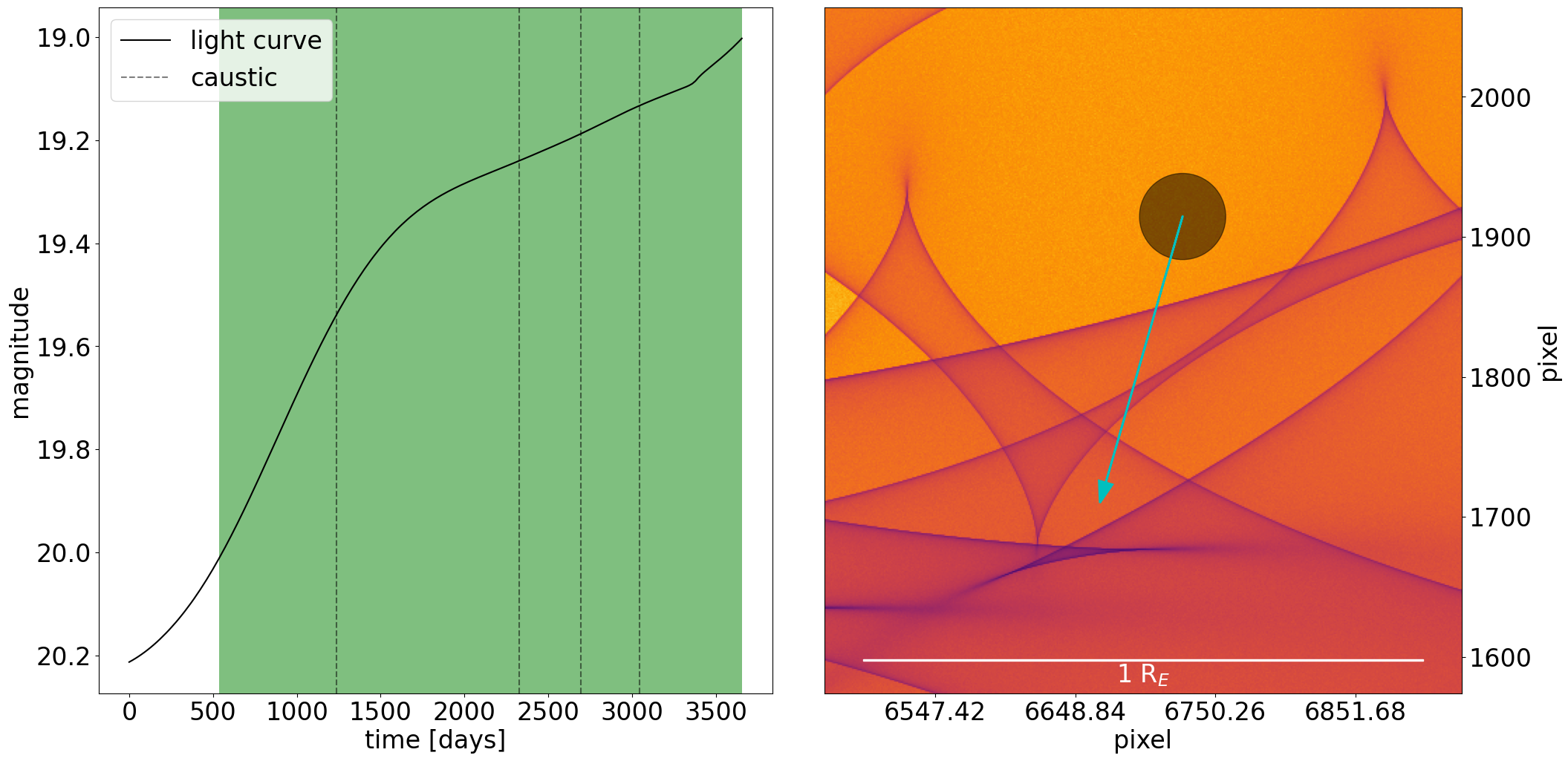}
     \caption{Same as Fig. \ref{fig:example_strong_weak}, but for a different system where the event is classified as a single caustic crossing event.
     We note that, in this example, a HME can be due to crossing multiple caustics, and its single classification is due to only identifying a single event.}
     \label{fig:example_single}
\end{figure*}

\begin{figure*}
     \includegraphics[width=1\columnwidth]{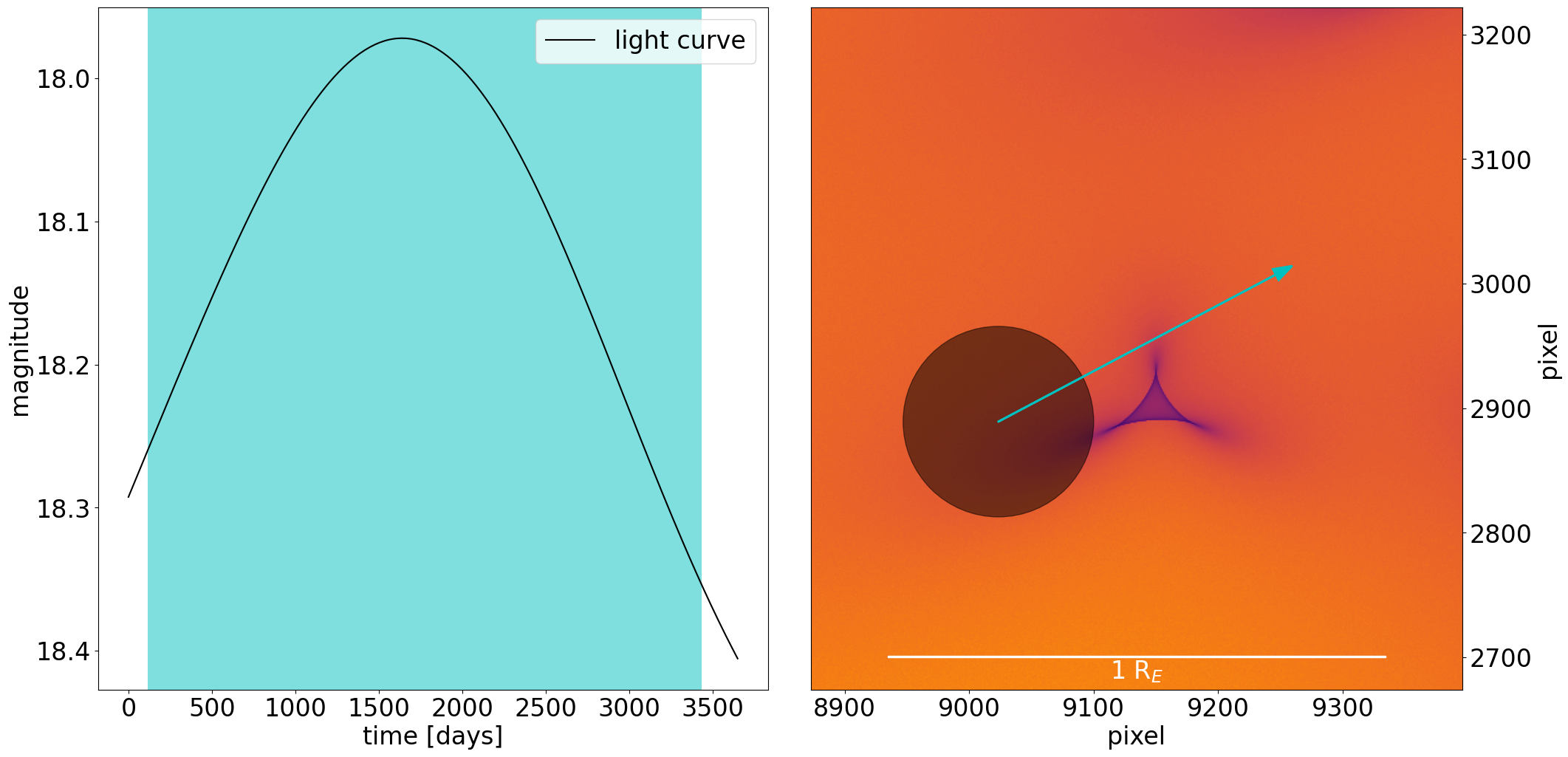}
     \caption{Same as Fig. \ref{fig:example_strong_weak}, but for a different system where instead we show two contiguous HMEs that meet at the brightness peak that are classified as non-caustic crossing and touching a caustic.}
     \label{fig:example_touching}
\end{figure*}

\begin{figure*}
     \includegraphics[width=1\columnwidth]{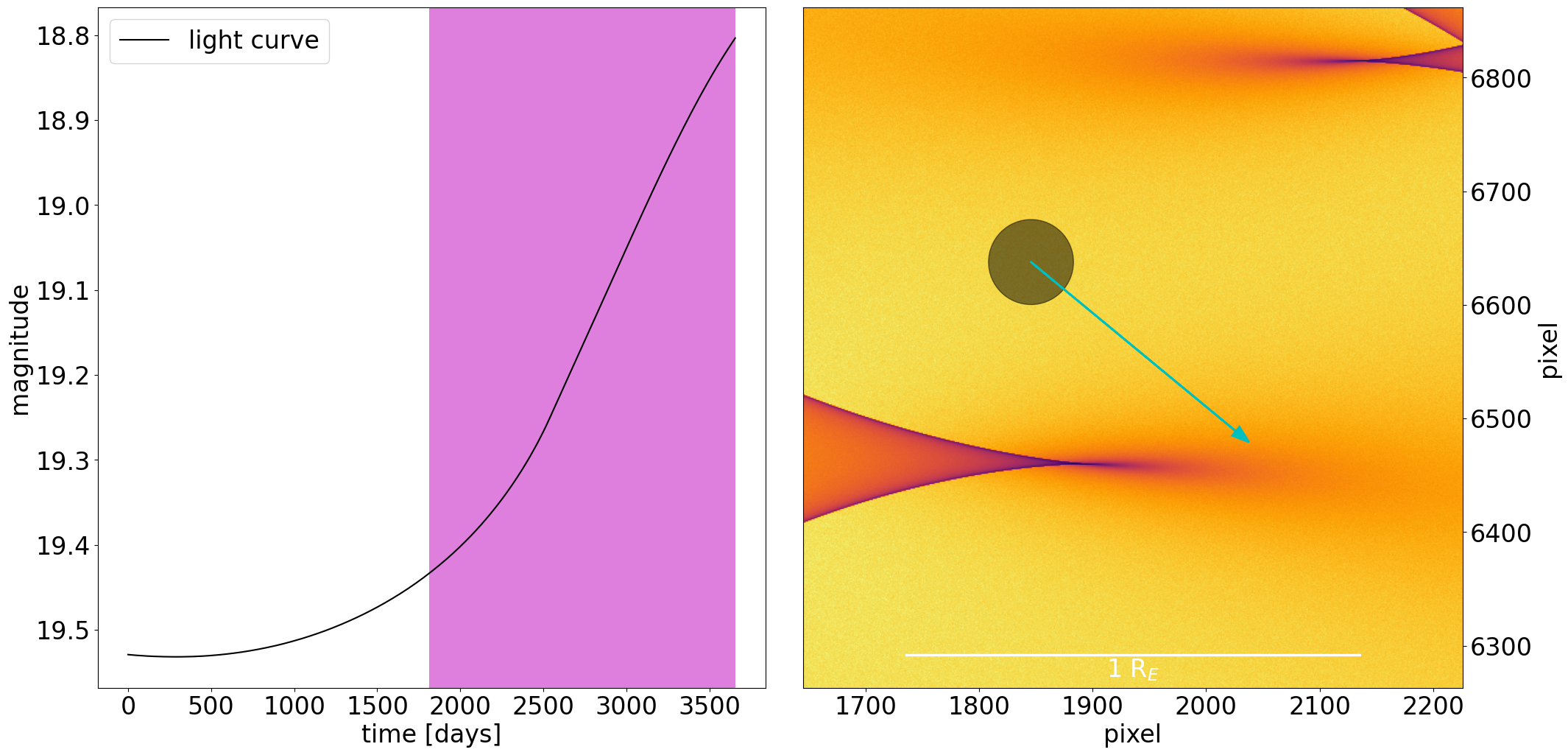}
     \caption{Same as Fig. \ref{fig:example_strong_weak}, but for a different system where we identify an event classified as non-caustic crossing and non-touching a caustic.}
     \label{fig:example_notouching}
\end{figure*}

In section \ref{sec:events} we presented a HME classification scheme (see Fig. \ref{fig:scheme}).
Here, from Fig. \ref{fig:example_strong_weak} through Fig. \ref{fig:example_notouching}, we show some examples of light curves where we have identified and highlighted their events and portray their corresponding classification.


\end{document}